\documentclass[12pt]{article}
\makeatletter
\newcommand{\lambdabar}{{\mathchoice
  {\smash@bar\textfont\displaystyle{0.25}{1.2}\lambda}
  {\smash@bar\textfont\textstyle{0.25}{1.2}\lambda}
  {\smash@bar\scriptfont\scriptstyle{0.25}{1.2}\lambda}
  {\smash@bar\scriptscriptfont\scriptscriptstyle{0.25}{1.2}\lambda}
}}
\newcommand{\smash@bar}[4]{%
  \smash{\rlap{\raisebox{-#3\fontdimen5#10}{$\m@th#2\mkern#4mu\mathchar'26$}}}%
}
\makeatother

\RequirePackage[a4paper,top=30.6mm,bottom=38.6mm,left=26mm,right=26mm,footskip=1.3cm]{geometry}
\usepackage{setspace}
\usepackage{amsmath}
\usepackage{amsfonts} 
\usepackage{amssymb}
\usepackage{graphicx}
\usepackage{hyperref}
\usepackage{tensor}
\usepackage{siunitx}
\usepackage{float}
\usepackage{bm}
\usepackage{slashed}
\usepackage{booktabs}
\usepackage{cancel}
\usepackage{multirow}
\usepackage{comment}
\usepackage{bbold}
\usepackage{caption}
\usepackage{amsmath}
\usepackage{enumerate}
\usepackage{cite}
\usepackage{tensor}
\usepackage{slashed}
\usepackage{inputenc}
\usepackage{rotating}
\usepackage{bigfoot}
\usepackage{cancel}
\usepackage{mathtools}
\usepackage{xcolor}
\usepackage{color, colortbl}

\usepackage[textsize=footnotesize]{todonotes}

\numberwithin{equation}{section}

\def \< {\left<}
\def \> {\right>}

\newcommand{\be}{\begin{equation}} \newcommand{\ee}{\end{equation}}
\newcommand{\bea}{\begin{eqnarray}}  \newcommand{\eea}{\end{eqnarray}}
\newcommand{\nn}{\nonumber}

\usepackage{amsfonts, amsthm}
\usepackage[english]{babel}
\usepackage{slashed}
\usepackage{mathrsfs}
\usepackage{amssymb}
\usepackage{color}

\begin{document}
		
	\begin{center}        % Main title 	
		\Large On the K-point in moduli space
		\end{center}
	
	\vspace{0.7cm}
	\begin{center}        % Authors
		{\large Jarod Hattab ,  Eran Palti , Joan Quirant}
	\end{center}
	
	\vspace{0.15cm}
	\begin{center}  
		\emph{Department of Physics, Ben-Gurion University of the Negev,}\\
		\emph{Be'er-Sheva 84105, Israel}\\[.3cm]
%		\emph{}\\[.2cm]

	\end{center}
	
	\vspace{1cm}
	
	%%%%%%%%%%%%%%%%%%%%%%%%%%%%%%%%%%%%%%%%%%%%%%%
	%%%%%%%%%%%%%%%%%%%%%%%%%%%%%%%%%%%%%%%%%%%%%%%
	%%%%%%%%%%%%%%%%%%%%%%%%%%%%%%%%%%%%%%%%%%%%%%%
	%%%%%%%%%%%%%%%%%%%%%%%%%%%%%%%%%%%%%%%%%%%%%%%
	%%%%%%%%%%%%%%%%%%%%%%%%%%%%%%%%%%%%%%%%%%%%%%%
	%%%%%%%%%%%%%%%%%%%%%%%%%%%%%%%%%%%%%%%%%%%%%%%
	%%%%%%%%%%%%%%%%%%%%%%%%%%%%%%%%%%%%%%%%%%%%%%%
	%%%%%%%%%%%%%%%%%%%%%%%%%%%%%%%%%%%%%%%%%%%%%%%
	
	\begin{abstract}
	\noindent  
    We study a class of infinite-distance loci, referred to as K-points, in one-parameter complex-structure moduli spaces of type IIB string theory compactified on Calabi-Yau manifolds. We show that around K-points the effective four-dimensional supergravity exhibits certain unusual properties. The two most prominent being that the leading order dependence of the prepotential on the gauge couplings is non-perturbative and that the leading gauge kinetic terms in the action vanish when evaluated on an anti self-dual graviphoton background. These properties are shared with the conifold locus in moduli space, rather than the large complex-structure locus. The conifold locus is well-known to arise from integrating out a charged BPS state, and so the similarities suggest that the K-point also arises from integrating out a BPS state. We develop such an interpretation, finding that it corresponds to a BPS state which is extremely light, whose mass in Planck units is doubly-exponentially small in the distance to the K-point. The state behaves as if it had complex charges, or as if it couples to the self-dual and anti self-dual parts of the graviphoton differently. Assuming such an integrating-out scenario is indeed the correct physics for the K-point, we discuss the implications for our understanding of infinite distances in moduli space and for the Swampland Distance Conjecture.
	\end{abstract}
	
	\thispagestyle{empty}
	\clearpage
	
	\tableofcontents

\section{Introduction}
\label{sec:int}

The Swampland program aims to determine universal properties of low-energy effective theories that arise from string theory \cite{Vafa:2005ui} (see \cite{Palti:2019pca,vanBeest:2021lhn} for reviews). One of the central themes is universal behaviour when approaching infinite distances in moduli space. The most well-known such proposal is the Distance Conjecture \cite{Ooguri:2006in}. This proposes that approaching any such infinite distance locus an infinite tower of states becomes exponentially (in the distance) lighter than the Planck scale. Evidence for the conjecture flows primarily from string theory tests, the best understood of which are in the context of extended supersymmetry. The setting with the richest set of examples, but which still maintains full control of the calculation, are four-dimensional effective theories with eight supercharges that arise from compactifications of type II string theory on Calabi-Yau manifolds. Initial investigations of this were in \cite{Palti:2017elp,Grimm:2018ohb}, and we refer to \cite{Palti:2019pca,vanBeest:2021lhn,Harlow:2022ich} for reviews and \cite{Cota:2022maf,Friedrich:2025gvs,Rudelius:2023odg,Gendler:2022ztv,Kaufmann:2024gqo,Marchesano:2022avb,Castellano:2022bvr,vandeHeisteeg:2022btw,Cribiori:2022nke,Marchesano:2022axe,Blumenhagen:2023tev,Castellano:2023qhp,Cribiori:2023ffn,Blumenhagen:2023yws,Blumenhagen:2023xmk,Seo:2023xsb,Calderon-Infante:2023ler,Calderon-Infante:2023uhz,Castellano:2023aum,Castellano:2023jjt,Castellano:2023stg,Cota:2023uir,Casas:2024ttx,Blumenhagen:2024ydy,Blumenhagen:2024lmo,Artime:2025egu,Hattab:2023moj,Hattab:2024thi,Hattab:2024chf,Hattab:2024yol,Hattab:2024ewk,Hattab:2024ssg,Monnee:2025ynn} for a small selection of more recent work.

The vector multiplet moduli space of the effective four-dimensional supergravity, which in the type IIB setting corresponds to the complex-structure moduli space of the Calabi-Yau, is very well controlled due to the extended supersymmetry. This means that it can probe very different regions in parameter space, where the light degrees of freedom vary significantly. Each region corresponds to an expansion about a singularity in moduli space. The most familiar is the large complex-structure region, which is mirror to the type IIA large volume region. Another familiar setting is the conifold region, where a three-cycle shrinks to zero size. A general classification of regions was initiated in \cite{Grimm:2018ohb,Grimm:2018cpv}, using the language of limiting mixed Hodge structures. 

The way that the theory probes the different regions is somewhat subtle: the theory is valid over the whole complex-structure moduli space only with a zero energy cutoff. At a given point in moduli space, there are a number of finite energy scales at which the effective theory breaks down. For example, it can be that the effective two-derivative four-dimensional description breaks down due to higher-derivative corrections, or due to light Kaluza-Klein modes. A universal sector which controls the cutoff of the theory is the presence of charged BPS states that have been integrated out in formulating the supergravity. That is, the supergravity is an effective description where in the path integral the charged BPS states have already been integrated over. The effective supergravity is therefore not valid whenever the degrees of freedom associated to these BPS states become dynamical. At any singularity in the moduli space, some BPS states become massless, and therefore the theory as defined over the whole moduli space breaks down at any finite energy scale. 

An effective theory which is valid at some non-zero energy scale corresponds to regions in the moduli space which are expanded about a singularity, but lie at some (finite or infinite) distance from it. It is therefore of interested to understand how the cutoff of the theory, as bounded by the BPS states dynamics, varies as a function of the distance to the singularity. The Distance Conjecture can be understood as describing this: for any infinite distance singularity, the cutoff decreases exponentially in the distance to the singularity. The cutoff also depends on other parameters of the theory, for example the value of the string coupling, but it is of most interest to us to determine its dependence on the complex-structure moduli space parameters.

% The cutoff set by the dynamics of BPS states is a scale where the degrees of freedom of the theory need to be rearranged. This is because the supergravity is formulated from the fundamental string perspective. On the other hand, the charged BPS states are associated to non-perturbative D-brane states, which are integrated out in the theory. They can appear as solitonic solutions, but not as fundamental dynamical degrees of freedom. There are therefore two expansions which control the effective theory: an expansion in moduli space about a singularity and an expansion in the energy cutoff scale set by the D-brane states dynamics. 

In this work we study one-parameter Calabi-Yau manifolds, so with a one (complex) dimensional moduli space. There are three regions in this space, expanded about three singular loci: the large complex-structure region, the conifold region, and the K-point region. The physics around the first two loci, the large complex-structure locus and the conifold locus, has been studied in detail. The third locus, the K-point, is an infinite distance type II$_0$ locus, in the language of \cite{Grimm:2018ohb,Grimm:2018cpv}. The physics of such points has been less studied. Earlier work on it is \cite{Grimm:2018ohb,Grimm:2018cpv,Joshi:2019nzi,Palti:2021ubp,Bastian:2023shf,erkinger2022spherepartitionfunctioncalabiyau,Knapp_2024,Erkinger_2023,Douaud:2024khu}. More recently, it was studied in \cite{Friedrich:2025gvs} in the context of Emergent String Limits (in the language of \cite{Lee:2019wij}) and duality with the Heterotic string. The picture from these works is that the physics of the K-point is much more similar to that of the large complex-structure locus than the conifold locus. This is expected because the former two are at infinite distance in moduli space. 

As discussed, the supergravity directly yields a zero energy description of the physics in the K-point region. There are a number of probes from the supergravity of finite scale physics. The one which has been mostly used so far, and was studied generally in \cite{Grimm:2018ohb}, is the mass spectrum of BPS states. This spectrum becomes light exponentially fast in the distance, suggesting that the theory has a weakly-coupled description up to that scale. However, there are also other probes, which have been investigated less. In this work we consider some of these, in particular the structure of the prepotential as a function of the gauge couplings and the form of the gauge kinetic matrix. 

We find that the K-point has the following properties. First, we show that the {\it leading order} dependence of the prepotential on the vector multiplet gauge coupling is exponential rather than polynomial. This is as opposed to the form one finds in the large complex-structure region where the leading dependence is polynomial, and there are exponential sub-leading corrections. Instead, it matches the form in the conifold region of moduli space. This suggests that its physics should bear similarities to the conifold locus. 

A second interesting feature of the K-point region is related to a special background of the supergravity which allows for non-vanishing constant field-strengths for the gauge fields, and which preserves the full ${\cal N}=2$ supersymmetry. This background is called the graviphoton background, because the field strengths are turned on only along the anti self-dual components of the graviphoton $W^-_{\mu\nu}$. This background has played a crucial role in our understanding of the relation between integrating out the non-perturbative BPS states and the supergravity effective theory from the fundamental string perspective, as developed in \cite{Bershadsky:1993cx,Antoniadis:1993ze,Gopakumar:1998jq,Dedushenko:2014nya,Hattab:2024ssg}. It turns out that at the K-point, the gauge fields kinetic terms in the action vanish at leading order when evaluated in this background. This is as opposed to the gauge kinetic terms in a generic background which diverge logarithmically when approaching the K-point. This property does not hold at the large complex-structure point, but does occur at the conifold point. So, again, we find similarities with the conifold locus, rather than the large complex-structure one. 

In this work we make the analogy between the K-point locus in moduli space and the conifold locus precise. In particular, it is known that the singular conifold locus has an interpretation in terms of integrating out a charged BPS state, which becomes massless on the locus \cite{Strominger:1995cz}. This is the analogue of the monopole locus in Seiberg-Witten theory \cite{Seiberg:1994rs}. We find that the K-point supports a similar interpretation as arising from integrating out a BPS state. However, crucially, the charges of this BPS state are required to be complex. Complex charges are a very exotic concept. However, recall that the state has already been integrated out of the theory, and so we are interpreting its contribution to the effective theory as if it had complex charges. Indeed, we show that an equal interpretation is that it has a type of chiral coupling to the graviphoton, coupling to its self-dual and anti self-dual parts differently. In any case, the results suggest that a theory which has the state as a dynamical degree of freedom would have to be quite exotic. We therefore find that the physics of the K-point may be much more exotic than previously thought. 

There is much that remains to be understood. The interpretation of the supergravity data in terms of finite energy physics, of BPS states and otherwise, is subtle. Certainly, an interpretation in terms of a state with complex charges warrants caution. We therefore do not claim that we have fully proven one picture or another of the K-point. Nonetheless, we believe that an interpretation in terms of integrating out a state should exist.  

Assuming that our description of the K-point region in terms of integrating out a BPS state is correct, we discuss its implications for the Swampland and the Distance Conjecture. First, it would be an example of an emergent infinite distance. The infinite distance singularity of the K-point would arise completely from integrating out the BPS state, and so occurs only as an infrared phenomenon. This aligns with the idea of the Emergence Proposal, which was stated in this context in \cite{Grimm:2018ohb}, related to ideas in \cite{Harlow:2015lma,Heidenreich:2017sim,Heidenreich:2018kpg}, and formulated generally in \cite{Palti:2019pca}. 

In terms of the Distance Conjecture, a crucial point is that the mass of the BPS state turns out to be doubly-exponentially small in the distance. That means that the effective theory breaks down far below the expected (single) exponentially low scale. The absence of a weakly-coupled description up to the exponentially low scale would violate certain formulations of the conjecture, and would only be consistent with a much less strong formulation (which we discuss).  

The paper is organized as follows. Section \ref{sec:eff} sets up the supergravity framework used in our analysis. Section \ref{sec:effregioncy} examines the supergravity locally around different regions in the moduli space. In section \ref{sec:gaugecoupre} we begin studying the physics around the K-point region by probing it with the dependence of the prepotential on the gauge couplings. In section \ref{sec:BPSint} we discuss the BPS states in the K-point region, and present an initial analysis of an integrating-out interpretation of the K-point. In section \ref{sec:elecmag} we introduce the graviphoton background, and discuss an interpretation of the K-point in terms of a background field integrating out calculation. In section \ref{sec:swdis} we discuss the implications of an integrating-out picture for the K-point for the Swampland Distance Conjecture. In section \ref{sec:disc} we summarise our results. Appendix \ref{sec:examcyx33} contains an explicit and detailed analysis of an example Calabi-Yau manifold. In appendix \ref{sec:revSW} we present a quick review of Seiberg-Witten theory.

\section{The supergravity framework}
\label{sec:eff}
 
We consider the setting of type IIB string theory compactified on a Calabi-Yau (CY) manifold to four dimensions. This leads to an ungauged ${\cal N}=2$ four-dimensional supergravity. This supergravity description is valid as long as the string coupling is weak, and all the Calabi-Yau cycles are large in string units. Our interest is in the vector multiplets moduli space, which is also the complex-structure moduli space of the Calabi-Yau. This moduli space exhibits certain singularities, about which it is possible to write local (relatively simple) effective theory descriptions. Because these effective theories need to fit into the general ${\cal N}=2$ supergravity framework, it is possible to write some general structures which are common to all of them. We discuss these in sections \ref{sec:genn2sug} and \ref{sec:wheregrav}. We then embed the effective theory on Calabi-Yau manifolds in this general formalism in section \ref{sec:effregioncy}. 

\subsection{General ${\cal N}=2$ supergravity formalism}
\label{sec:genn2sug}

For comprehensive reviews of ${\cal N}=2$ supergravity theories, see for example \cite{Andrianopoli:1996cm,Freedman_VanProeyen_2012}. We follow the conventions in \cite{Dedushenko:2014nya,Hattab:2024ssg}, and also those in the string compactifications literature, such as \cite{Gurrieri:2003st,Grimm:2005fa,Palti:2006yz,Palti:2008mg}. 

There is a gravitational multiplet whose bosonic fields are comprised of the graviton $h_{\mu\nu}$ and the graviphoton $V_{\mu}$. There are also $h^{(2,1)}$ vector multiplets, where $h^{(2,1)}$ is the Hodge number of the Calabi-Yau, labelled by an index  $i=1,...,h^{(2,1)}$. It is convenient to describe these multiplets with a redundancy as $h^{(2,1)}+1$ multiplets labelled by an index $I=0,i$. The bosonic components of these are labelled as $X_{(s)}^I$, and the vector bosons as $F^{I}_{\mu\nu}$. 

The redundancy in the description, with an additional vector multiplet, has two aspects to it. The first is related to the nature of the redundant gauge boson. This extra vector multiplet gauge field is identified with the graviphoton. This identification is quite subtle, and plays a very central role in this work. We therefore discuss it in depth in section \ref{sec:wheregrav}. 

We also have a redundancy in the description of the scalars in the vector multiplets. This corresponds to the gauge symmetry 
\be 
X_{(s)}^I \rightarrow \lambda X_{(s)}^I \;, 
\label{homresc}
\ee
for any $\lambda \in \mathbb{C}^*$. We can fix this gauge by writing
\be 
X_{(s)}^I = - \frac{i}{2} e^{\frac{K}{2}} X^I \;,
\label{xstox}
\ee 
and taking 
\be 
X^0 = 1 \;.
\label{x01gauge}
\ee 
In (\ref{xstox}) we introduced the real function $K$, which is the Kahler potential on the moduli space, defined in (\ref{kahlpot}). The relations (\ref{xstox}) define the $X^I$ coordinates which we use in this work.

%
% expression (\ref{kahlpot}) leads to two natural gauge choices for the rescaling parameter $\lambda$. The two choices as most natural in the context of using superfields and using only components of the superfields, and are\footnote{In the components gauge one usually also redefines the prepotential $F \rightarrow 2 F$, see \cite{Hattab:2024ssg} for a discussion.}
%\bea 
%\mathrm{Superfield\;gauge\;:\;} & &  X^0 = - \frac{i}{2} e^{\frac{K}{2}} \;, \nn \\
%\mathrm{Components\;gauge\;:\;} & & X^0  = \frac12 \;. 
%\eea 
%In this work we utilise the superfield gauge, but will connect to the components gauge often. We therefore define naturally coordinates $\Pi^I$ as
%\be 
%X^I = - \frac{i}{2} e^{\frac{K}{2}} \Pi^I \;, 
%\label{defPi}
%\ee 
%so that in the superfield gauge we have 
%\be 
%\Pi^0=1 \;.
%\label{gaugefixpi0}
%\ee 

The action for the vector multiplet fields takes the following form.\footnote{Our conventions are a mostly-plus signature for the metric $\left(-,+,+,+\right)$, and that $\epsilon_{0123}$ is positive.}
The kinetic terms for the vector fields are given by 
\be 
{\cal L}_{F} = \frac14\text{Im}\,\mathcal{N}_{IJ} F^{I}_{\mu\nu} F^{J,\mu\nu} -\frac18  \text{Re}\,\mathcal{N}_{IJ} F^{I}_{\mu\nu} F^{J}_{\rho\sigma} \epsilon^{\mu\nu\rho\sigma}\;,
\label{n2lfac}
\ee 
where 
\bea 
\mathcal{N}_{IJ}&=& \overline{F}_{IJ}+i\frac{N_{IL}X^{L} N_{JK}X^{K}}{N_{MN}X^{M}X^{N}} \;.
\eea 
Here we utilise the notation
\be 
F_{IJ}=\partial_{X^I}\partial_{X^J}F\left(X\right) \;,\;\; F_{I}=\partial_{X^I}F\left(X\right)\;,
\label{magper}
\ee 
and define
\be 
N_{IJ} = 2\;\text{Im}\,F_{IJ} \;.
\ee 
The (homogenous degree-two) function $F\left(X\right)$ is known as the supergravity prepotential. 
The matrix ${\cal N}$ relates electric and magnetic quantities, for example a useful relation is
\be 
F_I = {\cal N}_{IJ} X^J \;\;.
\ee

The physical moduli space is parameterised by gauge invariant combinations 
\be 
z^I = \frac{X^I}{X^0} \;.
\label{zidef}
\ee 
The associated kinetic terms in the action are then 
\be
{\cal L}_z = -g_{ij} \partial_{\mu}z^i\partial^{\mu}\overline{z}^j \;.
\label{supkint}
\ee 
The moduli space metric is given in terms of the non-homogenous coordinates as
\be 
g_{ij} = -\frac{\partial^2}{\partial z^{i}\partial \overline{z}^j}\log\left(-N_{KL}z^{K}\overline{z}^{L}\right) \;.
\label{Kahlmet}
\ee 
The metric on the moduli space $g_{ij}$ has an associated Kahler potential $K$, so $g_{ij} = \partial_i \bar{\partial}_j K$, given by 
\be
e^{-K} = \frac{i}{4\left|X^0\right|^2}\left(F_I \overline{X}^I - \overline{F}_I X^I \right)\;.
\label{kahlpot}
\ee
This completes the introduction of the general framework.

\subsection{The graviphoton}
\label{sec:wheregrav}

The four-dimensional graviphoton $V_{\mu}$ plays a central role in our analysis. It is defined as the vector superpartner to the graviton, so the vector in the gravitational multiplet. We denote its field-strength as $W_{\mu\nu}$, so
\be 
W_{\mu\nu} = \partial_{\mu} V_{\nu} - \partial_{\nu} V_{\mu} \;.
\ee 
The graviphoton field-strength is some combination of the vector field-strengths $F^I_{\mu\nu}$. However, this combination depends on the moduli, so the scalars in the vector multiplets. In order to determine this combination, we need to work with self-dual $F^+_{\mu\nu}$ and anti self-dual $F^-_{\mu\nu}$ field-strengths. The reason is that the supersymmetry spinors are chiral, and so couple to specific chirality combinations. The graviphoton is determines by its anti self-dual component $W^-_{\mu\nu}$. 

Let us define $F^{\pm}_{\mu\nu}$ as 
%\footnote{We are careful to state that these are the expression in Minkowski signature because in section \ref{sec:int} we discuss Wick rotating to Euclidean signature.}
\be 
\mathrm{Minkowski\;\;:\;\;} F^{I,\pm}_{\mu\nu} = \frac12 \left( F^I_{\mu\nu} \mp i\tilde{F}_{\mu\nu}^I\right) \;,
\label{fpmmink}
\ee 
with the magnetic field strength $\tilde{F}_{\mu\nu}^I$ defined as
\be 
\tilde{F}_{\mu\nu}^I = \frac{1}{2} \epsilon_{\mu\nu\rho\lambda} F^{I,\rho\lambda} \;.
\label{tildFidefm}
\ee 
We therefore have
\be 
F^I_{\mu\nu} = F^{I,+}_{\mu\nu} + F^{I,-}_{\mu\nu} \;,\;\; F^{I,\pm}_{\mu\nu} = \left( F^{I,\mp}_{\mu\nu}\right)^*\;.
\ee 
The anti-selfdual part of the 4d graviphoton field is then given by \cite{Freedman_VanProeyen_2012}
\be
W_{\mu\nu}^- = e^{\frac{K}{2}}\left(-X^I G_{I,\mu\nu} + F_I F^I_{\mu\nu} \right) = 2ie^{\frac{K}{2}}X^I\mathrm{Im\;}{\mathcal{N}}_{IJ}F^{J,-}_{\mu\nu}\;\;,
\label{gravWdef}
\ee
where the magnetic field-strengths are given by 
\be 
G_{I,\mu\nu}^- = \overline{{\cal N}}_{IJ}F^{J,-}_{\mu\nu} \;.
\label{GNFminrel}
\ee

The anti self-dual part of the graviphoton is related to the mass of BPS states, which is given by the integral of it over the sphere at infinity $S^2_{\infty}$. We can define the central charge $Z(q)$ as
\be 
Z(q)=-\frac{i}{2}\int_{S^2_{\infty}} W^-_{\mu\nu} \;dx^{\mu\nu} = -ie^{\frac{K}{2}}\left( q_I X^I - p^I F_I\right)  \;,
\label{defcenhargint} 
\ee 
where the quantized electric and magnetic charges are given by 
\be 
\int_{S^2_{\infty}} F^I_{\mu\nu} \;dx^{\mu\nu} = -2 q_I \;\;,\;\; \int_{S^2_{\infty}} G_{I,{\mu\nu}} \;dx^{\mu\nu} = -2 p^I \;\;.
\label{chargedefFint}
\ee 
The mass $M(q)$ of charged BPS states is then given by
\be 
M(q) = \left|Z(q) \right| \;.
\label{massBPSqg}
\ee 

Because the natural formulation of the graviphoton is in a self-dual basis, it is useful to write also the kinetic terms of the fields in this basis. The action (\ref{n2lfac}) can be written as
\be
{\cal L}_{F} = \frac{i}{4} \mathcal{\overline{N}}_{IJ} F^{I,-}_{\mu\nu} \left(F^{J,-}\right)^{\mu\nu}+ \mathrm{h.c.} \;\;\;.
\label{n2lfacasd}
\ee

\section{Effective descriptions in Calabi-Yau moduli spaces}
\label{sec:effregioncy}

In this section we apply the general supergravity framework to the moduli space of Calabi-Yau manifolds. The effective supergravity theory can be described locally around singularities in the vector-multiplet moduli space, which is the CY complex-structure moduli space. A general classification of these possible singularities, and the resulting effective theories, was presented in \cite{Grimm:2018ohb,Grimm:2018cpv}. In this work we focus on one parameter setups, so with a single complex-structure modulus. Studies of such models can be found, for example, in \cite{Joshi:2019nzi,Palti:2021ubp,Bastian:2023shf}. In particular, in \cite{Bastian:2023shf} a beautiful and detailed study of such models was presented, and we often utilise their results and conventions. 
%In this subsection we therefore will specialise, though still keeping some generality, the ${\cal N}=2$ framework to the notation and gauge choices of \cite{Bastian:2023shf}. 

\subsection{Local coordinates and gauge choices}
\label{sec:loccycochs}

Since we are restricting to one-parameter models, we denote the local (complex) coordinate in the moduli space as $s$, so that the singularity in the moduli space that we are expanding around is at $s=0$. The relation between $s$ and the vector-multiplet scalar components $X^I$ depends on the particular singularity. For one-parameter models, we have only $X^0$ and $X^1$. The homogenous gauge choice fixes $X^0=1$, and therefore we can determine the embedding by specifying the function $X^1(s)$.

It is also useful to describe the Kahler potential and prepotential in terms of the gauge-invariant coordinates $z^i$ in (\ref{zidef}). In one-parameter models there is only one such coordinate, which we denote as $z$. From the definition (\ref{zidef}) we see that
\be 
z = X^1(s) \;.
\label{zpis}
\ee  
In terms of this coordinate we can write the Kahler potential (\ref{kahlpot}) as 
\be
K(z) = - \log \left[ 2 i \left( F(z) - \overline{F}(\bar{z})\right) -i \left( z - \bar{z}\right) \left( \partial_z F(z) + \partial_{\bar{z}}\overline{F}(\bar{z}) \right) \right] + \log 4\;. 
\label{kahpotz}
\ee 

We are still required to specify the prepotential $F(z)$. We can do so locally around any singularity by expanding in $s$ around $s=0$. This takes the general form  \cite{Grimm:2018ohb}
\be
F(s) = \sum_{n=0}^3 f_n\left(s\right)\left( \log s \right)^n \;,
\label{expins}
\ee 
where the $f_n(s)$ are power series in $s$. 

The prepotential $F(z)$ is then determined by inverting the relation (\ref{zpis}), so solving for $s$ as a function of $z$, and then inserting the solution into the expansion (\ref{expins}). In general, the relation between $z$ and $s$ takes the form \cite{Grimm:2018ohb,Joshi:2019nzi,Palti:2021ubp,Bastian:2023shf}
\be 
z = \tau  +  \mu\; s + \frac{\lambda}{2\pi i}\; \log s  \;,
\label{zass}
\ee 
where $\tau$, $\mu$ and $\lambda$ are complex constants. 

There are three types of moduli space singularities in one-parameter CY models: the Large Complex-Structure (LCS) point, the conifold point, and the K-point.\footnote{These are the infinite order monodromy singularities. There are also finite order monodromy singularities, that are pure orbifold singularities. They are at finite distance with no massless states, and so are not of interest for this work.} In terms of the parameters in (\ref{zass}) we have that  
\bea 
\text{Large\;Complex-Structure}\;:\; & & \; \tau=\mu=0 \;,\;\; \lambda=1 \;, \nn \\
\mathrm{Conifold}\;:\; & & \;  \tau=\lambda=0 \;,\;\; \mu=1  \;,\nn \\
\mathrm{K\;point}\;:\; & & \;  \lambda=0 \;,\;\; \mu=1 \;.
\label{oneparasing}
\eea 
In terms of the distance to the singularities in moduli space, the LCS and K-points are at infinite distance, while the conifold is at finite distance. 

\subsection{The prepotential}
\label{sec:Prepo}
Having set up the $\mathcal{N}=2$ supergravity formalism and the evaluation of special geometry data in different regions of Calabi–Yau complex-structure moduli space, we now study and compare the behaviour of the holomorphic prepotential across the main boundary components: the large complex structure point, conifold points, and K-points. For each degeneration we use an adapted integral symplectic frame, which makes the universal singular pieces explicit and organises the rest into a regular power series. 
%In Section \ref{sec:Heterotic} we discuss a possible Heterotic dual description of the prepotential. 

\subsubsection{The large complex-structure region}
\label{sec:LCSPrepo}

The most familiar region of the theory is at large complex-structure. From (\ref{zass}), we have that
\be 
z =  \frac{1}{2\pi i}\log s  \;.
\ee 

Around the large complex-structure point there exists a natural choice of prepotential associated to a canonical integral frame. Although the integral frame is not unique and any $\mathrm{Sp}(4,\mathbb{Z})$ transformation acting on the period vector $(X^I,F_I)$ preserves the charge lattice, the large complex-structure frame is distinguished because the monodromy acts as a simple shift $z\mapsto z+1$ in the coordinate $z=X^{1}/X^{0}$. In this frame the holomorphic prepotential $F(z)$ makes the geometric data manifest 
\begin{equation}
\label{LCSPrepo} 
    F(z) = (2\pi i )^3\left(-\frac{\kappa}{6}z^3+\frac{a}{2}z^2+\frac{c}{24}z+\frac{\zeta(3)\chi}{2(2\pi i )^3}+\frac{1}{(2\pi i )^3}\sum_{\beta\geq1}\alpha_\beta\,\text{Li}_3(e^{2\pi i \beta z})\right)\;.
\end{equation}
Indeed, it is chosen such that under mirror symmetry, the prepotential matches the genus-zero A-model topological-string free energy on the mirror Calabi-Yau. Under mirror symmetry, the LCS frame matches directly to the type IIA large volume limit, where $z\mapsto z+1$ corresponds to the $B$-field shift and $(X^I,F_I)$ pairs naturally with the charges of D$(2p)$-branes (D0/D2,D4/D6). As such $\kappa, a, c,\chi$ and $\alpha_\beta$ are identified (up to some symplectic ambiguity) with enumerative properties of the mirror Calabi-Yau.

\subsubsection{The conifold region}
\label{sec:ConifoldPrepo}

In the conifold limit, from (\ref{zass}), we have that
\be 
z = s  \;.
\ee 
Then approaching the conifold locus $s \rightarrow 0$ we have 
\be 
-\log |s| \rightarrow \infty \;.
\ee

The large complex-structure frame is tailored to the maximally unipotent monodromy point and is not generally optimal for other boundary components of the Calabi-Yau moduli space. 
For a given degeneration, one prefers an adapted integral symplectic frame in which the monodromy and the singular part of the prepotential take a universal form. 
Such frames exist for ordinary conifold points and, more generally, for one-parameter degenerations \cite{Bastian:2023shf}. Near conifold singularities the universal local form reads
\be
\label{ConifoldPrepo}
F(s) = c_0-\frac{k}{4\pi i}s^2\log s+\frac{1}{2\pi i}\sum_{n\geq 1} b_ns^n\;,
\ee
where $k$ is an integer constant, $c_0$ a complex constant and the $b_n$ are rational numbers.\footnote{The integer $k$ has a geometric meaning : the conifold three-cycle is a sphere that is orbifolded by a factor $k$ \cite{Gopakumar:1997dv,Bastian:2023shf}}

\subsubsection{The K-point region}
\label{sec:KPrepo}

From \eqref{oneparasing}, a K-type one-parameter degeneration admits, in an adapted integral symplectic frame, a local coordinate \(s\) vanishing at the boundary such that
\begin{equation}
  z \;=\; \tau + s\,,
  \label{Kpoinzsre}
\end{equation}
with $\tau\in\mathbb{C}$ constant.

The parameter $\tau$ in (\ref{Kpoinzsre}) plays a central role in the physics of the K-point. It is possible to show with full generality \cite{Grimm:2018ohb,Bastian:2023shf} that $\tau$ satisfies an equation of the form
\be 
\left( \begin{array}{cc} 1 & \bar{\tau} \end{array} \right) {\bold B}\left( \begin{array}{c} 1 \\ \bar{\tau} \end{array} \right) =0\;\;, \;\;\;\;\; {\bold B} = \left( \begin{array}{cc} a & b \\ b & c \end{array} \right) \;\;,
\label{taugenr}
\ee 
with ${\bold B}$ an integer matrix, $a,b,c \in \mathbb{Z}$, with positive determinant 
\be
a c - b^2 > 0 \;.
\ee 
We can write it explicitly as
\be 
\tau = \frac{1}{c} \left( -b + i \sqrt{ac - b^2} \right) \;.
\label{tuexp}
\ee
Note that $|\tau|^2=\frac{a}{c}$. 
Some examples of the matrix ${\bold B}$ for different Calabi-Yau manifolds are \cite{Bastian:2023shf}
\be
{\bold B} = \left( \begin{array}{cc} 2 & 1 \\ 1 & 2 \end{array}\right) \;\;,\;\;
{\bold B} =  \left( \begin{array}{cc} 1 & 0 \\ 0 & 1 \end{array}\right) \;\;,\;\;
{\bold B} =  \left( \begin{array}{cc} 2 & 0 \\ 0 & 6 \end{array}\right) \;\;,\;\;
{\bold B} =  \left( \begin{array}{cc} 1 & 1 \\ 1 & 2 \end{array}\right) \;.
\ee
The first example is associated to the CY labelled as $X_{3,3}$, which was studied in detail also in \cite{Joshi:2019nzi,Palti:2021ubp}. In appendix \ref{sec:examcyx33}, we present a detailed analysis of this example. 

The K-point is defined as the $s \rightarrow 0$ limit, and so we have 
\be 
-\log |s| \rightarrow  \infty \;.
\ee 
As in sections~\ref{sec:LCSPrepo} and \ref{sec:ConifoldPrepo}, the holomorphic prepotential near \(s=0\) takes a universal form \cite{Grimm:2018ohb,Grimm:2018cpv,Joshi:2019nzi,Bastian:2023shf}
\be 
\label{KPrepo}
F(s) = h_0-\frac{c }{4\pi i}s\left(s+2i\text{Im}\,\tau\right) \log s+\frac{1}{2\pi i }\sum_{n\geq 1} g_n s^n \;.
\ee
Here \(h_0\in\mathbb{C}\) is a model-dependent complex constants, while the $g_n$ form a holomorphic power series with a certain arithmetic structure. 
In appendix \ref{sec:examcyx33}, we present a detailed analysis of an example CY where the prepotential (\ref{Fsx33}) and other relevant expressions are calculated from first principles.

Note that the structure of (\ref{KPrepo}) is much closer to the conifold case (\ref{ConifoldPrepo}) than to the large complex-structure one (\ref{LCSPrepo}), with the singular part involving only a linear logarithm contribution. 

\subsection{The gauge kinetic matrix}
\label{sec:gaukinmat}

In this section we calculate the gauge kinetic matrix, and Kahler potential, for the different regions in moduli space.

\subsubsection{The large complex-structure region}
\label{sec:LCSgauge}

We have that the large complex-structure point, where $s \rightarrow 0$, has $\mathrm{Im\;} z \rightarrow +\infty$, and so for simplicity we set 
\be 
\mathrm{Re\;}z =0 \;.
\label{lcsrez0r}
\ee 

The large complex-structure region has certain universal properties, which are well-known. In this section we utilize such general expressions, as calculated for example in \cite{Bastian:2023shf}. In appendix \ref{sec:examcyx33}, we present a detailed analysis of an example Calabi-Yau, that one can compare with. 
 
At large complex-structure, we have the expression for the Kahler potential
\be 
e^{-K} = \frac13 \kappa \left( \mathrm{Im\;} z \right)^3 + {\cal O}(1) \;.
\label{eKlcs}
\ee 
The integer $\kappa$ which appeared already in (\ref{LCSPrepo}) can be understood as the triple intersection number of the mirror Calabi-Yau. 
The gauge kinetic matrix behaves as \cite{Gurrieri:2003st,Grimm:2005fa,Bastian:2023shf} \footnote{We find a sign difference compared to the expression in \cite{Bastian:2023shf}. The same is true for (\ref{gkmKp}) and (\ref{tdef}). Details can be found in appendix \ref{sec:examcyx33}.} 
\be 
{\cal N} = -i\frac{\kappa}{6} \left( \begin{array}{cc} \left(\mathrm{Im\;} z \right)^2  & 0 \\ 0 &  3  \end{array} \right)\;\mathrm{Im\;} z + {\cal O}(1)\;.
\label{gkmlcsf}
\ee  
%\be 
%\mathrm{Im\;} {\cal N}_{IJ} = -\left( \begin{array}{cc} \left(\mathrm{Im\;} z \right)^2  & 0 \\ 0 &  3  \end{array} \right)_{IJ}\frac{\kappa}{6} \;\mathrm{Im\;} z + {\cal O}(1)\;.
%\label{gkmlcsf}
%\ee  

\subsubsection{The conifold region}
\label{sec:Conifoldgauge}

A crucial difference between the conifold and large complex-structure points is that in the conifold limit the Kahler potential remains finite
\be 
e^{-K} = \kappa_c + {\cal O}(s^2 \log s) \;,
\label{eKcon}
\ee 
where $\kappa_c$ is a constant.

The gauge kinetic function behaves as \cite{Bastian:2023shf}
\be 
{\cal N} = i\frac{k}{2\pi}\left( \begin{array}{cc} 0  & 0 \\ 0 & 1 \end{array} \right) \log |s|+ {\cal O}(1)\;,
\label{congkf}
\ee  
where $k$ is the integer constant introduced already in (\ref{ConifoldPrepo}).

\subsubsection{The K-point region}
\label{sec:Kgauge}

The  Kahler potential around a K-point is given by
\be 
e^{-K} = -\kappa_0 \log |s| + {\cal O}(s) \;,
\label{Kkpoi}
\ee 
%\be 
%\kappa_0 = \frac{4 c \left(\mathrm{Im}\; \tau \right)^2}{2\pi} \;.
%\ee 
with $\kappa_0$ a real positive constant. 
The diverging behaviour is similar to the large complex-structure point (\ref{eKlcs}), and in contrast to the conifold point (\ref{eKcon}), which can be understood as a consequence of the K-point being at infinite distance in moduli space. 

The gauge kinetic matrix reads at leading order \cite{Bastian:2023shf} 
\be   
{\cal N}_{IJ} = \frac{i}{2\pi} {\bold B}_{IJ} \log |s| + {\cal O}(1) \;,
\label{gkmKp}
\ee 
where ${\bold B}$ is the matrix in (\ref{taugenr}), comprised of real integer entries.

\section{Probing the theory with gauge couplings}
\label{sec:gaugecoupre}

Having gathered the appropriate data for the supergravity in the various regions in moduli space, in this section we begin the analysis of what this data can tell us about how the theory behaves in the different regions. In this section we consider the specific property of how the prepotential depends on the gauge couplings. We argue that this is an important diagnosis tool for the structure of the theory. We also, in section \ref{sec:het}, discuss the relation to duality with the Heterotic string around the K-point, and how the dependence on the gauge couplings manifests there. In section \ref{sec:KSW} we discuss a natural interpretation of the results of this section. 

\subsection{The large complex-structure region}
\label{sec:LCSprobe}

The gauge kinetic matrix at large complex-structure (\ref{gkmlcsf}), under the restriction (\ref{lcsrez0r}), factorises at leading order . Therefore, we can read off two gauge couplings. We choose to normalize the gauge couplings as
\be 
-\frac{1}{4g^2} F_{\mu\nu}F^{\mu\nu} \;,
\ee 
so that we have for the gauge couplings $g_0$ and $g_1$, associated to the field-strengths $F^0_{\mu\nu}$ and $F^1_{\mu\nu}$, the expressions
\be 
\left(g_0\right)^2 = \frac{6}{\kappa}\frac{1}{\left(\mathrm{Im\;}z\right)^3} + ...\;\;,\;\; \left(g_1\right)^2 =  \frac{2}{\kappa}\frac{1}{\mathrm{Im\;}z} + ...\;.
\ee 
It is informative to introduce complexified gauge couplings $S_I$ such that
\be 
\mathrm{Im\;}S_0 = \frac{4\pi}{\left(g_0\right)^2} \;\;,\;\;\mathrm{Im\;}S_1 = \frac{4\pi}{\left(g_1\right)^2} = 2 \pi \kappa \;\mathrm{Im\;}z + ... \;\;.
\ee 
We can then write the prepotential (\ref{LCSPrepo}) in terms of the gauge couplings. There is some freedom in doing so because there are two gauge couplings and only one modulus $z$. In any case, we are only interested in the qualitative dependence on the gauge couplings and so write
\begin{equation}
\label{LCSPrepogauge}
    F(S_1) = \frac{i}{6 \kappa^2} \left(S_1\right)^3  + ... + {\cal O}\left( e^{\frac{iS_1}{\kappa}} \right) \;.
\end{equation}
%\begin{equation}
%\label{LCSPrepogauge}
%    F(g) =  \frac{(2\pi)^3}{\left(g_0\right)^2} + \text{polynomial in $g$} + {\cal O}\left( e^{-\frac{4\pi}{k\left(g_1\right)^2}} \right) \;.
%\end{equation}
The general structure is that the leading dependence is polynomial in the (complexified) gauge couplings, and then there are terms that are exponential in the inverse gauge couplings.  

The exponential terms are not arising from spacetime instantons, but  are classical from the fundamental type IIB string perspective. They do have an interpretation as worldsheet instantons in the type IIA dual, and as spacetime instantons in a Heterotic dual. They also have an interpretation in terms of integrating out BPS states associated to wrapped D-branes \cite{Gopakumar:1998jq}. 

\subsection{The conifold region}
\label{sec:Conifoldprobe}

In the conifold region, the leading behaviour of the gauge kinetic matrix is divergent along one direction (\ref{congkf}). This singles out a gauge coupling
\be 
\left(g_1\right)^2 = -\frac{2\pi}{k} \frac{1}{\log |s|} + ... \;.
\label{g1conlime}
\ee 
Let us define the phase of $s$ as 
\be 
s = |s| e^{i \theta_c} \;,
\ee 
then we can write the holomorphic completion of (\ref{g1conlime}) in terms of a complexified gauge coupling $S_c$ as
\be 
S_c = \theta_c + \frac{4 \pi i}{\left(g_1\right)^2} = -2ki\log s + ... \;.
\ee

The prepotential in the conifold region (\ref{ConifoldPrepo}) can then be written as
\be
\label{ConifoldPrepogauge}
F(S_c) = c_0- \frac{1}{8\pi}\; S_c \;e^{\frac{iS_c}{k}} +...\;.
\ee
In contrast to the large complex-structure gauge coupling dependence (\ref{LCSPrepogauge}), in the conifold region the leading order dependence on the (complexified) gauge coupling is exponential, rather then polynomial.

\subsection{The K-point region}
\label{sec:Kprobe}

In the K-point region, the leading order gauge coupling (\ref{gkmKp}) is not diagonal. However, since the matrix ${\bold B}$ is just a simple matrix with order one integer entries, its diagonalization does not alter much the structure of the gauge couplings. We have two gauge couplings which have, up to order one coefficients, the same dependence on the parameter $s$. We can therefore write both of their complexified gauge couplings as $S_K$ with
\be 
S_K = - 2 f i \log s + ...\;,
\label{Skdef}
\ee  
with $f$ some order one coefficent which would be different for the two gauge couplings (and is simple to calculate). 

The leading dependence of the prepotential around the K-point (\ref{KPrepo}) on the gauge couplings can then be written as
\be
\label{KPrepogauge}
F(S_K) = h_0+ \frac{c \,\mathrm{Im\,}\tau}{4\pi i f}\; S_K \;e^{\frac{iS_K}{2f}}  + ... \;.
\ee
The dependence is exponential, as in the conifold region (\ref{ConifoldPrepogauge}), and in contrast to the large complex-structure region (\ref{LCSPrepogauge}). 

This is the first signal that the K-point region is somewhat exotic. The form of the prepotential suggests that it shares properties of the conifold region, and possibly there is some interpretation in terms of integrating out a BPS state. On the other hand, the K-point is at infinite distance in moduli space, as opposed to the conifold locus which is at finite distance. 

The K-point in complex-structure moduli space is a special case of type II loci, in the notation of \cite{Grimm:2018cpv}. In \cite{Grimm:2018cpv}, and further work, up to the most recent \cite{Friedrich:2025gvs}, the K-point was considered similar in nature to the large complex-structure locus. In the rest of this paper we study the possibility that it shares much with the conifold locus, including whether we can understand its leading gauge coupling behaviour as arising from integrating out a light BPS state. 

\subsection{A Heterotic perspective}
\label{sec:het}

In \cite{Friedrich:2025gvs} it was proposed that the K-point region has a weakly-coupled Heterotic dual description. The Heterotic string has a target space of the form $K3 \times T^2$. The appearance of a $K3$ was discussed already in \cite{Grimm:2018ohb}, further in \cite{Friedrich:2025gvs}, and mathematically in \cite{tyurin2003fanoversuscalabi,doran2016mirrorsymmetrytyurindegenerations,Calabrese_2015}.\footnote{In fact, the $K3$ is why it is called the K-point \cite{vanstraten2017calabiyauoperators}.}

The formulation of the prepotential in terms of the gauge couplings is well adapted to describing a possible Heterotic dual. This is because the gauge couplings are controlled by the (complexified) Heterotic string coupling $S_H$. We can write this as
\be 
S_H = a_H + \frac{4\pi i}{\left(g_H\right)^2}   \;,
\label{hetaxdil}
\ee 
where $a_H$ is the Heterotic universal axion and $g_H$ is the (four-dimensional) Heterotic string coupling. 

By matching BPS states around the K-point to KK and Winding modes of the Heterotic string, a weakly-coupled Heterotic dual description is one where we identify in the K-point region \cite{Friedrich:2025gvs}
\be 
\text{Weakly coupled dual\;:}\;\;\; S_H = - 2 i f_H  \log s + ... \;,
\label{hetKdual}
\ee 
with $f_H$ some constant.

Using the relation (\ref{hetKdual}), we therefore have the prepotential taking the form
\be
\label{HetPrepogauge}
F(S_H) = h_0+ \frac{c\,\mathrm{Im\,}\tau}{4\pi i f_H}\; S_H \;e^{\frac{iS_H}{2f_H}}  + ... \;.
\ee
We find that the dependence on the dilaton $S_H$, or Heterotic string coupling $g_H$, is completely non-perturbative. This is in contrast to what happens at something like the weakly-coupled setting where the $K3 \times T^2$ is in the geometric regime. There the dilaton appears linearly in the prepotential, in the famous $STU$ combination. This difference is a Heterotic manifestation of the IIB gauge coupling dependence. It suggests that a Heterotic dual would not be weakly-coupled, at least not up to the string scale, as in a perturbative Heterotic string scenario. 

\subsection{A Seiberg-Witten analogy}
\label{sec:KSW}

The conifold region in moduli space is closely related to the famous Seiberg-Witten description of $SU(2)$ gauge theory physics \cite{Seiberg:1994rs}. This was first pointed out in \cite{Strominger:1995cz}, and later developed significantly. We refer to \cite{Lerche:1996xu} for an excellent review. The similarities between the conifold and K-point prepotential dependence on the gauge couplings can be framed within this context. In this section we briefly develop this interpretation. We include a brief review of Seiberg-Witten theory in appendix \ref{sec:revSW}, so that in this section the idea can be more succinctly stated. 

In $SU(2)$ gauge theory, there is a (complexified) dynamical scale $\Lambda$, below which the theory is strongly-coupled. Near the origin of the Coulomb branch, where the gauge group is broken $SU(2) \rightarrow U(1)$, there is a description of the strongly-coupled theory in terms of a weakly-coupled magnetic frame theory. This magnetic infrared theory has a vector multiplet moduli space parameterised in terms of a (dual) variable $a_D$. The magnetic $U(1)$ theory is controlled by a (complexified) weak magnetic coupling $S_D$, which takes the form (\ref{mcosw})
\be 
S_D =\frac{1}{2\pi i} \log \left( \frac{a_D}{\Lambda} \right)+ ... \;.
\label{mcoswd}
\ee
The coupling is weak as long as 
\be 
\left| \frac{a_D}{\Lambda} \right| \ll 1 \;.
\ee 
The prepotential in the magnetic theory $F\left(S_D\right)$ takes the form (\ref{Fswadtemy})
\be 
F\left(a_D\right) = \frac{1}{4\pi i} a_D^2 \log \left( \frac{a_D}{\Lambda} \right)+ ... \;.
\label{Fswadtemy}
\ee 
Then if we write it in terms of $S_D$, we have the exponential dependence
\be 
F\left(S_D\right) = \frac{1}{2} \;S_D \;e^{4 \pi i S_D}+ ... \;.
\label{expFdSWext}
\ee 
This exponential dependence is of similar form to (\ref{ConifoldPrepogauge}) and (\ref{KPrepogauge}). 

In the Seiberg-Witten description, the exponential form of the prepotential (\ref{expFdSWext}) is well understood. It can be traced back to the gauge coupling form (\ref{mcoswd}), which arises from integrating out a BPS monopole state, with mass $M_{\mathrm{mon}}$, given by
\be 
M_{\mathrm{mon}} =\left|a_D\right| \;.
\ee 

The analogy with the conifold region is therefore clear: the exponential prepotential dependence on the gauge coupling is a manifestation of the fact that the leading order behaviour of the gauge coupling (\ref{congkf}) is induced from integrating out a BPS state. 

We can phrase this relation more generally, as follows. The infrared value of the gauge coupling in the conifold region, $g_1$ in (\ref{g1conlime}), is controlled by the the vector multiplet modulus through the mass of an integrated out BPS state. So (\ref{g1conlime}) is a threshold correction to some ultraviolet gauge coupling, which dominates in the infrared. The ultraviolet coupling is not sensitive to the modulus $s$, the analogue of $a_D$, and does not diverge as $s \rightarrow 0$. Similarly, the leading term in the prepotential is not sensitive to $s$, and appears as a constant. The leading term which depends on $s$ then arises in the infrared as a threshold correction, which in the prepotential then appears exponentially in the gauge coupling.  

It is therefore natural to propose a similar interpretation for the K-point region. In such an interpretation the dependence on $s$ of the prepotential and gauge coupling is an infrared one, induced by integrating out a BPS state below a dynamical scale. So, in this scenario, varying $s$ around the K-point is not a direct variation of some ultraviolet gauge coupling but an indirect variation of an infrared gauge coupling induced by a threshold correction. 

It is worth noting a simple point, which holds for both the conifold and K-point interpretations. In the Seiberg-Witten setting, the theory with the gauge coupling (\ref{mcoswd}) is only valid in the infrared, below the mass of the monopole. That is, the running of the coupling is replaced by the threshold correction appearing in (\ref{mcoswd}). Above the monopole mass scale, we have to reinstate the scale dependence coming from integrating out momentum modes of the monopole state. Therefore, if we define our infrared theory as one where the monopole has been integrated out, we have a cutoff on this theory of  
\be 
\Lambda_{\mathrm{IR}} \sim |a_D| \;. 
\label{lambdair}
\ee 
In the Seiberg-Witten case, it is possible to integrate the monopole state back into the theory. So we can work with a theory above the scale $\Lambda_{\mathrm{IR}}$ , but below the dynamical scale $\Lambda$. In the string theory setting this is not possible. There is no known gravitational description where the conifold state is dynamical. Similarly, in a conifold-like interpretation of the K-point region, there would be no local gravitational description above the mass of the integrated-out state. So in string theory, $\Lambda_{\mathrm{IR}}$ is really the cutoff of any (known) local gravitational description. 

\section{BPS states and integrating out}
\label{sec:BPSint}

The large complex-structure prepotential (\ref{LCSPrepo}), and the conifold prepotential (\ref{ConifoldPrepo}) both have well-known relations to the spectrum of BPS states in the theory. 

In the case of the conifold, the leading dependence of the prepotential on the gauge coupling can be understood in terms of integrating out a single BPS state, corresponding to a D3 brane wrapping a three-cycle, which becomes massless on the conifold locus \cite{Strominger:1995cz}. 

In the large complex-structure region, the exponential terms in the prepotential can also be understood as arising from integrating out BPS states \cite{Gopakumar:1998jq}. The leading terms, polynomial in $z$, can also be connected to integrating out BPS states, though in a more subtle way. This connection was developed in \cite{Hattab:2023moj,Hattab:2024chf,Hattab:2024ewk,Hattab:2024ssg,Hattab:2024thi} and in \cite{Blumenhagen:2023tev,Blumenhagen:2023xmk,Blumenhagen:2023yws,Blumenhagen:2024lmo,Blumenhagen:2024ydy,Blumenhagen:2025zgf}.

In this section we study whether the leading K-point prepotential, or more precisely the gauge kinetic matrix, can be understood as arising from integrating out a BPS state. Our focus will be on an interpretation similar to that of the conifold locus. 

%The latter is known to be closely related to Seiberg-Witten theory, and so for completeness we present a very quick review of the relevant physics in section \ref{sec:revSW}. 
%

%\subsection{The conifold BPS state}

\subsection{BPS states around the K-point}

BPS states associated to wrapped D-branes are characterised by real integral charges $q_I$ and $p^I$ as in (\ref{defcenhargint}). Given the charges, the central charge is defined in (\ref{chargedefFint}), which we reproduce here for convenience
\be 
Z\left(q\right) = -ie^{\frac{K}{2}}\left( q_I X^I - p^I F_I\right)  \;. 
\label{simcentchar}
\ee 
The mass $M(q)$ of BPS states is then fixed by supersymmetry as the absolute value of the central charge (\ref{massBPSqg}). The phase of the central charge is mapped to the particular ${\cal N}=1$ supersymmetry preserved by the (half) BPS state within the full ${\cal N}=2$. 

The quantization of the charges $q_I$ and $p^I$ is fixed at large complex-structure. It is then maintained in the K-point region because the period vectors in the two regions are related by an integral $Sp\left(4,\mathbb{Z}\right)$ transformation.

The period vector around the K-point $\Pi = (X^0,X^1,F_0,F_1)$ has the  following behaviour (see \cite{Grimm:2018cpv,Grimm:2018ohb,Joshi:2019nzi,Bastian:2023shf} or (\ref{expanprepo}) for a particular example)
\begin{equation}
    \Pi(s) = \left(\begin{array}{c} X^0 \\ X^1 \\ F_0 \\ F_1 \end{array}\right)= \left(\begin{array}{cccc}
 1\\
 \tau+s \\
\delta +\gamma\tau +\tau \alpha\log(s\beta)+\mathcal{O}(s\log(s))\\\
\gamma-\alpha\log(s\beta)+\mathcal{O}(s\log(s))
\end{array}\right)\;,
\end{equation}
where $\delta$ and $\gamma$ are rational, while $\alpha$ and $\beta$ are complex constants. The central charge (\ref{simcentchar}) therefore takes the form
\be 
Z\left(q\right) = -ie^{\frac{K}{2}}\;\Big[\;\left( q_0+q_1\tau - p^0(\delta+\gamma\tau)-p^1\gamma\right) - \alpha\left( p^0 \tau - p^1 \right)\log(s\beta) +\mathcal{O}(s\log(s))\Big] \;. 
\label{simcentchar}
\ee 

For arbitrary $\tau$, $\gamma$ and $\delta$ we therefore have three type of states whose mass, in Planck units, vanishes at the K-point $|s| \rightarrow 0$\;:
\bea 
p^0 \tau - p^1 = 0  \;\;,\;\; q_0+q_1\tau - p^0(\delta+\gamma\tau)-p^1\gamma \neq 0 \;\;&:&\;\; M \sim \frac{1}{\sqrt{-\log |s|}} \;\;, \nn \\
p^0 \tau - p^1 = 0 \;\;,\;\; q_0+q_1\tau - p^0(\delta+\gamma\tau)-p^1\gamma = 0 \;\;&:&\;\; M \sim |s| \sqrt{-\log |s|} \;\;, \nn \\
p^0 =p^1 = 0 \;\;,\;\; q_0+q_1\tau  = 0 \;\;&:&\;\; M \sim \frac{|s|}{\sqrt{-\log |s|}} \;\;.
\label{zstatekgen}
\eea 

Because $\tau$ is complex (\ref{tuexp}), for integral charges only the first type of states, whose mass behaves as $\left(-\log |s|\right)^{-\frac12}$, are possible. Such states are labelled by two integers, and we refer to them as Type G states (following the notation of \cite{Grimm:2018ohb,Grimm:2018cpv,Palti:2019pca})
\be 
\text{Type G} \;\;:\;\; p^0=p^1=0 \;\;,\;\; q_0\;, q_1 \in \mathbb{Z} \;\;,\;\; M \sim \frac{1}{\sqrt{-\log |s|}} \sim  \frac{1}{\sqrt{\left|S_K\right|}}\;.
\label{integstat}
\ee 
These states are the ones that have been extensively studied and discussed in the literature, for example in \cite{Grimm:2018cpv,Grimm:2018ohb,Joshi:2019nzi,Bastian:2023shf,Friedrich:2025gvs}. As discussed in section \ref{sec:swdis}, such states become light at an exponential rate in the distance to the K-point, and play the role of the charged tower in the distance conjecture. 

The tower of type G states around the K-point is similar in nature to the tower of  BPS states around large complex-structure. They become light as a power of the gauge coupling, so with mass scaling as $\left| S_K \right|^{-\frac12}$ with $S_K$ defined in (\ref{Skdef}). On the other hand, around the conifold, the light BPS state has a mass which is exponential in the gauge coupling 
\be 
\text{Conifold}\;\;:\;\;M_c \sim |s| \sim \left|e^{\frac{iS_c}{2k}}\right|\;.
\ee
Such an exponentially light state is required in order to interpret the leading behaviour of the gauge coupling as arising from integrating it out. 

Exponentially light states, like the BPS state at the conifold, are termed type F states in \cite{Grimm:2018ohb,Grimm:2018cpv,Palti:2019pca}. At the K-point, the (lightest) type F state would be one where
\be 
\text{(Lightest) Type F} \;\;:\;\; p^0=p^1=0 \;\;,\;\; q_0+q_1\bar{\tau}=0 \;\;,\;\; M \sim \frac{|s|}{\sqrt{-\log |s|}}  \sim \frac{1}{\sqrt{\left|S_K\right|}} \left|e^{\frac{iS_K}{2f}}\right|\;.
\label{typeFkstate}
\ee
Note that here we work with the conjugate conditions to (\ref{zstatekgen}), that is, we work with $\bar{Z}(q)$ rather than $Z(q)$. The reason for this choice is discussed in section \ref{sec:elecmag}. A conifold-type interpretation for the K-point requires the presence of such a state, or multiple such states. As discussed in section \ref{sec:gaugecoupre}, this is what the form of the prepotential suggests. On the other hand, charges $q_I$ which satisfy (\ref{typeFkstate}) must be complex
\be 
q_0+q_1\bar{\tau}=0 \implies q_0,q_1 \in \mathbb{C} \;.
\label{gyoncondit}
\ee 
The solution, with convenient normalization, to the constraint (\ref{gyoncondit}) is
\be 
\left( \begin{array}{c} q_0 \\ q_1 \end{array} \right) = \sqrt{c}\;\left( \begin{array}{c} -\bar{\tau} \\ 1 \end{array} \right) \;.
\label{gyoncharges}
\ee 
We therefore have a tension: at the K-point the prepotential is exponential at leading order in the gauge couplings, which suggests the it arises from integrating out a state satisfying (\ref{gyoncondit}). On the other hand, we do not expect complex charges to be possible. 

In this work we consider the setup where this tension is resolved by allowing for a state with complex charges. We discuss the possible meaning of this in section \ref{sec:elecmag}. First, we consider whether taking the K-point as arising from integrating out such a state can make sense. 

\subsection{An initial look at integrating out a complex-charged state}
\label{sec:intcomchss}

In this section we consider the possible existence of a state which behaves as effectively having the charges (\ref{gyoncharges}). Of course, all the charged BPS states have already been integrated out of the supergravity theory, and so this state is not dynamical. However, we are interested in this section in understanding to what extent the supergravity around the K-point behaves as if a state of such complex charges was integrated out. The analysis is a preliminary one, with a more precise framework developed in section \ref{sec:elecmag}.

\subsubsection*{The gauge kinetic matrix}

We first consider the gauge kinetic matrix (\ref{gkmKp}). 
Integrating the state with complex charges (\ref{gyoncharges}) at one loop can be expected to yield a contribution to the gauge kinetic matrix of the form
\be 
\left.\mathrm{Im\;} {\cal N}\right|^{1-\mathrm{loop}} \sim \left( \begin{array}{cc} \left|q_0\right|^2  & q_0 \bar{q}_1 \\ q_1 \bar{q}_0 & \left|q_1\right|^2 \end{array} \right) \log |s| \sim  \left( \begin{array}{cc} a  & b+i\sqrt{ac - b^2} \\ b-i\sqrt{ac - b^2} & c \end{array} \right) \log |s| \;.
\label{gkmintcc}
\ee 
When contracting (\ref{gkmintcc}) with the gauge field strengths, we find
\be 
\left.\mathrm{Im\;} {\cal N}_{IJ}\right|^{1-\mathrm{loop}} \;F^{I}_{\mu\nu} F^{J,{\mu\nu}} \sim  {\bold B}_{IJ} \;\log |s|\;F^{I}_{\mu\nu}  F^{J,{\mu\nu}}  \;,
\ee 
which is indeed the form of the kinetic term in the supergravity action (\ref{gkmKp}).   The point being that the complex parts of (\ref{gkmintcc}) disappear in the action due to the symmetric properties of the kinetic terms. 

Another way to view that the K-point singularity is associated to a single state of complex charge comes from the monodromy about that locus in moduli space. This monodromy is associated to the Witten effect where magnetically charged particles can pick up the charge of the massless particle. For the K-point, the monodromy matrix ${\bold T}$ acting on the set of electric $e_I$ and magnetic $p^I$ charges, forming a charge vector ${\bf e}$, is \cite{Bastian:2023shf}
\be 
{\bold T} = e^{\bold N} \;\;,\;\;\; {\bold N} = -\left( \begin{array}{cc} 0  & 0 \\ {\bold B} & 0 \end{array} \right) \;\;,\;\;\; {\bf e} = \left( \begin{array}{c} p^I   \\  e_I \end{array} \right) \;.
\label{tdef}
\ee
Under such a monodromy transformation, the electric charges $e_I$ of a general state will pick up a contribution from the magnetic charges $p^I$ of the form\footnote{The monodromy acts on the period vector as ${\bold T}$, and on the charges as ${\bold T}^{-1}$.}
\be 
e_I \rightarrow e_I + {\bold B}_{IJ}\; p^J \;. 
\label{monB}
\ee 
On the other hand, if an electric particle of charge $q_I$ is massless, it should induce a monodromy of the form
\be 
e_I \rightarrow e_I + q_J\;q_I\; p^J  \;. 
\label{monpq}
\ee 
So we would like to match (\ref{monB}) with (\ref{monpq}). However, it is simple to check that this is not possible for real charges with a single state $q_I$. This follows from the fact that the determinant of ${\bold B}$ is non-vanishing
\be
\det {\bold B} \neq 0 \;, 
\ee
which means that the transformation (\ref{monB}) will involve two free charge parameters, while in (\ref{monpq}) we only have one combination appearing: $q_J\; p^J$. 

However, we saw that integrating out the complex charged state actually leads to a matrix (\ref{gkmintcc}), and so it is natural to consider the monodromy action as
\be 
\widetilde{{\bold B}} =  \left( \begin{array}{cc} a  & b+i\sqrt{ac - b^2} \\ b-i\sqrt{ac - b^2} & c \end{array} \right) = \left( \begin{array}{cc} a  & - c \;\bar{\tau} \\ 
- c \;\tau & c \end{array} \right) \;.
\ee 
This matrix has a vanishing determinant 
\be 
\det \widetilde{{\bold B}} = 0 \;,
\ee
which means that if we replace ${\bold B}$ with $\widetilde{{\bold B}}$ we should be able to write (\ref{monB}) in the form (\ref{monpq}). More precisely, since we are dealing with complex charge, we should consider the transformation 
\be 
e_I \rightarrow e_I + \bar{q}_J\;q_I\; p^J  \;. 
\label{monpqc}
\ee
Now we can see the match between (\ref{monpqc}) and (\ref{monB}) (with $\widetilde{{\bold B}}$)  directly, since as in (\ref{gkmintcc}), we have 
\be 
\widetilde{{\bold B}}_{IJ} =  \bar{q}_J\;q_I \;.
\ee 

\subsubsection*{The prepotential}

A naive guess at the effect of integrating out the state on the gauge kinetic function (\ref{gkmintcc}), seems to yield sensible results. Within the supergravity framework, we can connect the leading behaviour of the gauge kinetic function with the prepotential. Indeed, in the conifold region, we can integrate out the charged BPS state into the gauge kinetic function or directly into the prepotential. 

At the K-point the relation is more subtle. This is because a prepotential integrating out calculation for a BPS state of mass $|s|$, should yield a prepotential contribution of the form $s^2 \log s$. See, for example \cite{Dedushenko:2014nya,Hattab:2024ssg}, for such a general supersymmetric integrating out procedure. This contribution can be seen for the conifold in (\ref{ConifoldPrepo}). On the other hand, in the K-point region, the leading $s$ dependence of the prepotential comes from a term behaving as $s \log s$ (\ref{KPrepo}).  We explain how to understand this difference in section \ref{sec:elecmag}.

\subsubsection*{A dyonic-like state}

There is another way to view the charges of the state, which suggest indeed non-locality and a relation to Argyres-Douglas theories \cite{Argyres:1995jj}. 

For standard real integer charges $q_I$, using (\ref{fpmmink}), we can write 
\be 
q_I F^I_{\mu\nu} = 2\;\mathrm{Re} \left( q_I F^{I,-}_{\mu\nu} \right) \;.
\label{qfstar1}
\ee 
We cannot do this for complex charges $q_I$. However, we could consider taking the right-hand-side of (\ref{qfstar1}) as our starting point, and complexify the charges in it. In other words, in the standard effective supergravity with real charges (\ref{qfstar1}) holds as an equality, so we can use either form. If we want to understand the meaning of complex charges, there is then some ambiguity as to which form in (\ref{qfstar1}) we should use. With respect to this ambiguity, we may as well consider using directly the right-hand-side of (\ref{qfstar1}). In section \ref{sec:charparbackw}, we explain how (\ref{qfstar1}) can be related to a background field integrating-out calculation. 

Since the $F^{I,-}_{\mu\nu}$ are complex, we can perform a change of basis to $\hat{F}^{I,-}_{\mu\nu}$ by a complex matrix ${\bf R}$ as
\be 
\hat{F}^{I,-}_{\mu\nu} =  {\bf R}^I_{\;\;K}\; F^{K,-}_{\mu\nu} \;.
\ee 
We can then write the combination $q_I F^{I,-}_{\mu\nu}$ appearing in (\ref{qfstar1}) in terms of charges $\hat{q}_I$ that couple to the $\hat{F}^{I,-}_{\mu\nu}$ basis as
\be 
q_I F^{I,-}_{\mu\nu} = q_I \left({\bf R}^{-1}\right)^I_{\;\;J} {\bf R}^J_{\;\;K}\; F^{K,-}_{\mu\nu}
 =   \left(\left({\bf R}^{-1}\right)^T\right)^{\;\;I}_{J} q_I \hat{F}^{J,-}_{\mu\nu} = \hat{q}_J\hat{F}^{J,-}_{\mu\nu} \;.
\ee
Choosing the form
\be 
{\bf R} = \left( \begin{array}{cc} 1 & 0 \\ -\sqrt{c}\;\bar{\tau}  & \sqrt{c}\end{array} \right)  \;,\;\; \left({\bf R}^{-1}\right)^T = \left( \begin{array}{cc} 1 & \bar{\tau} \\ 0  & \frac{1}{\sqrt{c}} \end{array} \right) \;,
\ee 
we have
\be 
 \left( \begin{array}{c} \hat{q}_0 \\ \hat{q}_1  \end{array} \right) =  \left( \begin{array}{c} 0 \\ 1 \end{array} \right)\;\;\;,\;\;\; \left( \begin{array}{c} \hat{F}^{0,-} \\ \hat{F}^{1,-}  \end{array} \right) =  \left( \begin{array}{c} F^{0,-} \\ \sqrt{c} \;F^{1,-} - \sqrt{c}\;\bar{\tau}\; F^{0,-} \end{array} \right) \;.
 \label{gyoncharotha}
\ee  

From (\ref{gyoncharotha}) we see that we can write the right-hand-side of (\ref{qfstar1}) as 
\be 
2\;\mathrm{Re} \left( q_I F^{I,-}_{\mu\nu} \right) = 2\;\mathrm{Re} \left( \hat{F}^{1,-}_{\mu\nu} \right)= \sqrt{c}\; \Big[ F^1_{\mu\nu} - \left( \mathrm{Re\;}\tau \;F^{0}_{\mu\nu} + \mathrm{Im\;}\tau \;\tilde{F}^{0}_{\mu\nu}\ \right) \Big] \;.
\label{rotfhqfi}
\ee 
We therefore see that if we consider using the right-hand-side of (\ref{qfstar1}) in the theory, rather than the left-hand-side, the complex charges of the state become just the unit charge under $\hat{F}^{1,-}_{\mu\nu}$, and if we read off what would be the electric field strength associated to $\hat{F}^{1,-}_{\mu\nu}$, we arrive at the combination in (\ref{rotfhqfi}). This combination includes both $F^{0}_{\mu\nu}$ and $\tilde{F}^{0}_{\mu\nu}$ simultaneously. Further, the relative factors are irrational, which means that there is no electromagnetic transformation which would make the combination purely electric or purely magnetic. In this sense, it seems that the state should be considered as a type of dyon. 

Having dyonic states appear in moduli space, and some non-locality being associated to it, is not so unusual in itself. Argyres-Douglas theories have this property \cite{Argyres:1995jj}. However, in such theories one has multiple dyons which are not mutually local becoming massless. In this setting, we have only one such dyon. In some sense, we could then consider the K-point as an isolated Argyres-Douglas type point, where only one state becomes massless. Indeed, the one-parameter Calabi-Yau in which the K-point appears can be understood as an orbifolding of multiple parameter Calabi-Yau manifolds. Therefore, we should be able to construct the isolated K-point as an orbifolding of a multi-parameter locus in complex-structure moduli space. This suggests that the theory there could be an orbifolding of an Argyres-Douglas type theory, and that this state could be related to some orbifold-invariant combination of mutually non-local dyons.

The analysis here was based on applying the relation (\ref{qfstar1}). In order to understand better its possible meaning, and more generally how the state should be integrated out, we need to introduce the concept of the graviphoton background. We go on to do this in section \ref{sec:elecmag}.

%\subsubsection*{A single-chirality state?}
%
%The mapping to a dyonic state relies on the relation (\ref{qfstar1}), and in particular in using the right-hand-side with complex charges. There is a natural idea as to what this could mean : we could consider the possibility that the state only has one chirality, and so couples only to the self-dual part of the gauge field. In that case we could write
%\be 
%q_I F^{I}_{\mu\nu} = q_I F^{I,+}_{\mu\nu} + q_I F^{I,-}_{\mu\nu} \rightarrow q_I F^{I,-}_{\mu\nu} = \hat{q}_I \hat{F}^{I,-}
%\ee 
%In this way the charge of the state is simply the unit charge, and it is charged only with respect to $\hat{F}^{1,-}_{\mu\nu}$. This is the same as what happens in the conifold case. So it could possibly be thought of as a single chirality version of the conifold. 
%
%Note that here the single-chirality would have to include the anti-particle, so violate CPT. It would be a truncation similar to self-dual Yang-Mills or self-dual gravity theories. These are not unitary local theories, but they seem consistent in many ways. 
%
%Indeed, (anti) self-dual fields play a direct role in understanding the relation between integrating out BPS states and the resulting effective supergravity. This is particularly so for the graviphoton. We go on to discuss this in section \ref{sec:elecmag}.

\section{Integrating out in the graviphoton background}
\label{sec:elecmag}

All the charged BPS states have been integrated out into the complex-structure, or vector-multiplets, moduli space. This is because all such states are non-perturbative from the fundamental string perspective. Therefore, an understanding of this integrating out is very subtle, because the theory in which the integrating out calculation could be performed is not one defined by a fundamental string. This is true even for the conifold locus, where the relation to the integrating out picture is perhaps simplest. As opposed to Seiberg-Witten theory, where the monopole state can be integrated in and out in the framework of a quantum field theory, there is no such framework for describing the conifold state dynamically. The same is also true at large complex-structure. There, there exists an M-theory dual where the pure D0 states can be treated as fundamental degrees of freedom in a (five-dimensional) supergravity framework, but the D2-D0 bound states are never fundamental and always have a super-Planckian mass within a supergravity regime (in terms of the higher-dimensional Planck scale).\footnote{This can be seen simply by noting that they are uplifted in M-theory to M2 branes, whose tension is set by the eleven-dimensional Planck scale, that wrap two-cycles, which must be large in those same Planck units in a supergravity regime.} 

Nonetheless, there is a quantitative integrating out calculation that can be performed, say at large complex-structure, even though the states are super-Planckian in the supergravity regime. This calculation is a background field calculation in the worldline formalism. It was first performed by Gopakumar and Vafa \cite{Gopakumar:1998jq}, and we refer to \cite{Dedushenko:2014nya,Hattab:2024ssg} for further recent developments and a review. The calculation, as developed in detail in \cite{Dedushenko:2014nya,Hattab:2024ssg}, is performed in Euclidean space with a constant field-strength for the anti self-dual component of the graviphoton. As discussed below, the latter is required because the calculation is performed on-shell, so with a background which solves the equations of motion. 
%This is quite a crucial point: the worldline for the particle that is integrated out is calculated in an on-shell background. 

The calculation is such that we consider turning on a spatially constant background for the gauge fields, which we denote as $ \left<F^I_{\mu\nu} \right>$ with
\be
\label{backFint}
\partial_{\rho} \left<F^I_{\mu\nu} \right> = 0\;.
\ee 
We can then trace over fluctuations in the path integral about the saddle corresponding to this background, and determine the contribution to the effective action. As emphasised in \cite{Dedushenko:2014nya}, it is crucial for this approach that the background is on-shell, that is, it solves the equations of motion, so is a saddle of the (Euclidean) action. 

Apart from solving the (Euclidean) equations of motion, the graviphoton background has a second remarkable property: it preserves the full ${\cal N}=2$ supersymmetry \cite{Ooguri:2003qp,Berkovits:2003kj,Dedushenko:2014nya}. Usually, we would expect a background to preserve only one half of the supercharges, so ${\cal N}=1$. The extended supersymmetry of the graviphoton background is crucial because a given charged BPS state, that is to be integrated out, will preserve only ${\cal N}=1$ supersymmetry. The particular combination of supercharges that is preserved is fixed by the phase of the central charge. That means that if we try to integrate out BPS states which preserve different ${\cal N}=1$ sub-groups of the supersymmetry, so with central charges that have different phases, the resulting effective action will not be an ${\cal N}=2$ F-term. Explicitly, each BPS state will have associated to it Goldstino modes that are the Goldstone modes of the broken supercharges. These Goldstino modes become the superspace parameters of the effective action. Writing an effective action as an F-term means it is an integral over half of superspace, so with four supersymmetry parameters. This implies that the BPS states must all preserve four unbroken supercharges. But if the background itself breaks four supercharges, then only a special sub-set of BPS states will preserve the same supercharges as the background. Specifically, BPS states which have different central charge phases, cannot simultaneously preserve the same supercharges as an ${\cal N}=1$ background. The importance of the graviphoton background preserving ${\cal N}=2$ is that any BPS state will preserve ${\cal N}=1$ and so, by an appropriate rotation, can all be integrated out simultaneously into an F-term. In turn, it is known that the two derivative ${\cal N}=2$ supergravity action is precisely an F-term. So any background-field interpretation of integrating out BPS states of the effective action terms needs an ${\cal N}=2$ background, so the graviphoton background.

The graviphoton background leads to a certain fundamental tension between real values for the gauge fields and supersymmetry. There is no fully understood resolution to this tension, even at large complex-structure. At the K-point, this issue becomes even more involved, and the complex charges that were considered in section \ref{sec:BPSint} are a manifestation of this tension. We go on to describe this tension and its implications.

\subsection{The anti self-dual graviphoton background}
\label{sec:ansfgbacks}

There is an important projection of the $F^{I,-}_{\mu\nu}$ field-strengths onto the graviphoton given by \cite{Freedman_VanProeyen_2012,Dedushenko:2014nya}
\be 
F^{I,-}_{\mu\nu}=\frac{i}{4} e^{\frac{K}{2}}\overline{X}^{I}W^-_{\mu\nu} \;.
\label{onshelcondgen}
\ee 
This relation plays a central role in our analysis. A background expectation value for the gauge field-strengths (\ref{backFint}) must satisfy (\ref{onshelcondgen}) to preserve supersymmetry. This follows from the appearance of the combination in the ${\cal N}=2$ vector superfields \cite{Freedman_VanProeyen_2012,Dedushenko:2014nya}.  We refer to such a background, of constant field-strengths satisfying  (\ref{onshelcondgen}) as the {\it graviphoton background}, and denote it as $\left<W^-_{\mu\nu} \right>$. The graviphoton background plays a special role in the relation between BPS states and the effective supergravity action \cite{Bershadsky:1993cx,Antoniadis:1993ze,Gopakumar:1998jq,Gauntlett:2002nw,Ooguri:2003qp,Berkovits:2003kj,Dedushenko:2014nya,Hattab:2024ssg}.

In Lorentzian signature, we can write the graviphoton background as
\be 
\text{Lorentzian\;\;:\;\;}\left<F^{I,-}_{\mu\nu}\right>=\frac{i}{4} e^{\frac{K}{2}}\overline{X}^{I}\left<W^{-}_{\mu\nu} \right> \;, \; \left<F^{I,+}_{\mu\nu}\right>=-\frac{i}{4} e^{\frac{K}{2}}X^{I}\left<W^{+}_{\mu\nu} \right>  \;.
\label{onshelcondL}
\ee 
The form of the background (\ref{onshelcondL}) is required from supersymmetry. On the other hand, there are two more constraints coming from the equations of motion for the scalar fields and for the metric. Keeping a flat background metric requires that the background is purely anti self-dual
\be 
\text{Flat spacetime} \;\;:\;\; \left<F^{I,+}_{\mu\nu}\right> = 0\;.
\label{flatspcons}
\ee 
Sourcing no scalar field gradients requires that it is solely along the graviphoton direction (since the graviphoton mutliplet has no scalar fields). The central point is that in a Lorentzian spacetime we have the relation
\be 
\text{Lorentzian\;\;:\;\;} \left<F^{I,-}_{\mu\nu}\right> = \left<F^{I,+}_{\mu\nu}\right>^* \;.
\ee 
Therefore, it is not possible to restrict the background to only anti self-dual components, as in (\ref{flatspcons}), while keeping the gauge field-strengths real. 

In non-supersymmetric settings, we face similar issues with instanton calculations. The answer there, as here, is to Wick rotate to Euclidean space. 
In Euclidean space, the relation (\ref{fpmmink}) becomes
\be 
\mathrm{Euclidean\;\;:\;\;} F^{I,\pm}_{\mu\nu} = \frac12 \left( F^I_{\mu\nu} \pm \tilde{F}_{\mu\nu}^I\right) \;.
\label{fpmeuc}
\ee 
This means that it is possible to treat $F^{I,\pm}_{\mu\nu}$ as independent, while keeping the gauge field-strengths real. So in Euclidean space, it is possible to take a background with
\be 
\mathrm{Euclidean\;\;:\;\;} \left<F^{I,-}_{\mu\nu}\right> \neq 0 \;\;,\;\; \left<F^{I,+}_{\mu\nu}\right>=0 \;\;,\;\; \;\;F^I_{\mu\nu} \;, \;\tilde{F}^I_{\mu\nu} \;, \;F^{I,\pm}_{\mu\nu} \in \mathbb{R} \;.
\ee 

However, now the tension with supersymmetry arises. The Euclidean space version of the graviphoton background (\ref{onshelcondL}) that we want to impose is 
\be 
\mathrm{Euclidean\;\;:\;\;}\left<F^{I,-}_{\mu\nu}\right>=\frac{i}{4} e^{\frac{K}{2}}\overline{X}^{I}\left<W^-_{\mu\nu} \right> \;,\;\; \left<F^{I,+}_{\mu\nu}\right>=0\;.
\label{onshelcondE}
\ee 
But in Euclidean space the $\left<F^{I,-}_{\mu\nu}\right>$ are real, while the right-hand-side of (\ref{onshelcondE}) is complex. So there is a fundamental tension which does not allow us to keep all three of : supersymmetry, flat spacetime, and real gauge fields. 

\subsection{Kinetic terms in the graviphoton background}

Because the integrating out calculation is performed in the graviphoton background, evaluating the action on the background yields valuable information on the BPS states that have been integrated out. 

We consider the Lorentzian action, and evaluate it on the background (\ref{onshelcondL}). Using the expressions in section \ref{sec:effregioncy}, the graviphoton background reads (at leading order)
\bea 
\label{gravbacklimits}
%\text{Large complex-structure}\;\;&:&\;\; \left<F^{I,-}_{\mu\nu}\right> =  \left( \begin{array}{c} i \\ \;\mathrm{Im\;} z  \end{array} \right)_I \frac{1}{4} \left(\frac{3}{\kappa}\frac{1}{\left(\mathrm{Im\;}z\right)^3} \right)^{\frac12} \left< W^-_{\mu\nu} \right> + ...\;, \\
\text{Large complex-structure}\;\;&:&\;\; \left<F^{I,-}_{\mu\nu}\right> =  \left( \begin{array}{c} i \\ \;\mathrm{Im\;} z  \end{array} \right)_I \sqrt{\frac{3}{16\kappa}} \left(\mathrm{Im\;}z\right)^{-\frac32}  \left< W^-_{\mu\nu} \right> + ...\;, \nn \\
\text{Conifold}\;\;&:&\;\;\left<F^{I,-}_{\mu\nu}\right>  =  \left( \begin{array}{c} 1 \\ \bar{s}  \end{array} \right)_I \frac{i}{4\sqrt{\kappa_c}}  \left< W^-_{\mu\nu} \right> + ... \;,  \\
\text{K-point}\;\;&:&\;\; \left<F^{I,-}_{\mu\nu}\right>  =  \left( \begin{array}{c} 1 \\ \; \bar{\tau} + \bar{s} \end{array} \right)_I \frac{i}{4\sqrt{-\kappa_0\log |s|}}   \left< W^-_{\mu\nu} \right> + ... \nn\;.
\eea

The gauge kinetic terms (\ref{n2lfacasd}) evaluated on the background (\ref{onshelcondL}) read 
\be
{\cal L}_{\left<W\right>} = -\frac{i}{64} e^K \left(\mathcal{N}_{IJ} X^I X^J \right)^* \left<W^{-}_{\mu\nu}\right>\left< W^{-,\mu\nu}\right>  + \mathrm{h.c.}\;\;.
\label{n2lfacasdgp}
\ee
%Because the divergences in $\mathcal{N}_{IJ}$ are in the imaginary part, we have that in the different regions, approaching $s \rightarrow 0$, the kinetic terms (\ref{n2lfacasdgp}) behave to leading order as
%\be 
%{\cal L}_{\left<W\right>} =  -\frac{1}{64} e^K \mathrm{Im\;}\mathcal{N}_{IJ} \overline{X}^I \overline{X}^J  \left<W^{-}_{\mu\nu}\right>\left< W^{-,\mu\nu}\right>+ ... + \mathrm{h.c.}\;\;.
%\label{n2lfacasdgpI}
%\ee 
For the different regions we have
\bea 
\label{kinonW}
\text{LCS}\;\;&:&\;\; {\cal L}_{\left<W\right>} = -\frac{1}{64} \left<W^{-}_{\mu\nu}\right>\left< W^{-,\mu\nu}\right>+ \mathrm{h.c.} +...  \;,  \\
\text{Conifold}\;\;&:&\;\;{\cal L}_{\left<W\right>} = \left(   k\underbrace{\bar{s}^2\; \log |s|}_{\rightarrow 0} + {\cal O}\left(1\right)  \right)\; \frac{\left<W^{-}_{\mu\nu}\right>\left< W^{-,\mu\nu}\right>}{128 \pi  \kappa_c}+ \mathrm{h.c.} \;, \nn \\
\text{K-point}\;\;&:&\;\; {\cal L}_{\left<W\right>} =\left[ -\underbrace{\left( \begin{array}{cc} 1 & \bar{\tau} \end{array} \right) {\bold B}\left( \begin{array}{c} 1 \\ \bar{\tau} \end{array} \right)}_{=0} \log |s| + {\cal O}\left(1\right)\right] \frac{\left<W^{-}_{\mu\nu}\right>\left< W^{-,\mu\nu}\right>}{-128 \pi  \kappa_0\log |s|}+ \mathrm{h.c.} \;. \nn
\eea
We therefore see another qualitative difference between the K-point and large complex-structure, and similarity between the K-point and the conifold: the contribution of the leading divergent part of the gauge kinetic matrix to the kinetic terms (behaving as $\log |s|$) vanishes when evaluated on the graviphoton background.
%It is informative to write the tension without direct reference to the graviphoton. By using the expression for the graviphoton (\ref{gravWdef}), we can write (\ref{onshelcondL}), or (\ref{onshelcondE}), as 
%\be 
%\left<F^{I,-}_{\mu\nu}\right>=
%\frac{i}{4} e^{K}\overline{X}^{I}\left(-X^J \left<G_{J,\mu\nu}\right> + F_J \left<F^J_{\mu\nu} \right>\right)   
%%= -\frac{1}{2} e^{K}\overline{X}^{I} X^J\mathrm{Im\;}{\mathcal{N}}_{JK}\left<F^{K,-}_{\mu\nu}\right>
%\;.
%\label{onshelcondWdef}
%\ee 
%We can then evaluate this in the large complex-structure, conifold and K-point limits in moduli space. Using the expressions in section \ref{sec:effregioncy}, we find
%\bea 
%\text{Large complex-structure}\;\;&:&\;\; -\frac12 \;\overline{X}^{I} \;, \nn \\
%\text{Conifold}\;\;&:&\;\;-\frac12 \; \overline{X}^{I} \;, \nn \\
%\text{K-point}\;\;&:&\;\; -\frac12 \; \overline{X}^{I} \;.
%\eea

\subsection{Charged particles in the background}
\label{sec:charparbackw}

The vanishing contribution of the leading (divergent) parts of the gauge kinetic matrix in (\ref{kinonW}), which occurs for the conifold and the K-point, suggests that they have some similarities. In the case of the conifold, this vanishing is attributed to the nature, and specifically charge, of the BPS state that is integrated out. Specifically, at the conifold the charges of the integrated out state are
\be 
\text{Conifold}\;\;:\;\;\left( \begin{array}{c} q_0 \\ q_1  \end{array} \right) =  \left( \begin{array}{c} 0 \\ 1 \end{array} \right) \;.
\label{conchargvc}
\ee
From (\ref{gravbacklimits}) we see that asymptotically at the conifold the graviphoton background aligns with $\left<F^{0,-}_{\mu\nu}\right>$. Therefore, the conifold state is uncharged under the asymptotic graviphoton. 
We can write this (at leading order in $s$) as
\bea 
\text{Conifold, Lorentzian}\;:\;q_I \left<F^{I}_{\mu\nu}\right> 
=\frac{i}{4\sqrt{\kappa_c}} \bar{s} \left< W^-_{\mu\nu} \right> - \frac{i}{4\sqrt{\kappa_c}} s \left< W^+_{\mu\nu} \right> + ...\;,
\label{concouback}
\eea 
which vanishes as $s \rightarrow 0$. 

We now consider the K-point. The starting point is the Euclidean background (\ref{onshelcondE}). We can go to basis where the graviphoton background matches the conifold form. This is nothing but the basis (\ref{gyoncharotha}), which gives a graviphoton background of 
\bea 
\text{K-point, Euclidean}\;\;:\;\;\left<\hat{F}^{I,-}_{\mu\nu}\right>  &=& \frac{i}{4\sqrt{-\kappa_0\log |s|}}   \left( \begin{array}{c} 1 \\ \sqrt{c} \;\bar{s}  \end{array} \right)_I \left< W^-_{\mu\nu} \right> + ...\;\;, \nn \\
\left<\hat{F}^{I,+}_{\mu\nu}\right> &=& 0 \;.
 \label{gyoncharothapb}
\eea  
The change of basis to (\ref{gyoncharotha}) is the natural one because of the vanishing of the self-dual parts of the fields in the background, which means we can write 
\be 
\text{K-point, Euclidean}\;\;:\;\;q_I \left<F^{I}_{\mu\nu}\right> = q_I \left<F^{I,-}_{\mu\nu}\right> = \hat{q}_I\left<\hat{F}^{I,-}_{\mu\nu}\right> \;.
\label{chbafpp}
\ee 
The background (\ref{gyoncharothapb}) matches, up to a normlization factor and at leading order in $s$, the form of the graviphoton background around the conifold locus (\ref{gravbacklimits}). The charges (\ref{gyoncharotha}) also match that of the conifold state
\be 
 \text{K-point, Euclidean}\;\;:\;\;\left( \begin{array}{c} \hat{q}_0 \\ \hat{q}_1  \end{array} \right)_{\left< W_{\mu\nu} \right>} =  \left( \begin{array}{c} 0 \\ 1 \end{array} \right)_{\left< W_{\mu\nu} \right>} \;,
 \label{kpochabpeuc}
 \ee 
 where the subscript $\left< W_{\mu\nu} \right>$ reminds us that these are charges in terms of how the state couples to the background field-strengths, so as in (\ref{chbafpp}). 
 Therefore, in the Euclidean background field integrating out calculation, there is no difference between the conifold and K-point BPS state with respect to the how the charges couple to the background. The K-point state behaves, with respect to that background, as a  unit-charged state! 
 
 Another way to think of this is that in the (Euclidean) graviphoton background the relation (\ref{qfstar1}) can hold even for complex $q_I$. More precisely, we have that\footnote{Note that for complex charges $q_I$ we take, as in (\ref{gyoncondit}), the $q_I$ as appearing in $\overline{Z}(q)$. So $Z(q)$ would then have the $\bar{q}_I$ appearing.}
\be 
q_I \left<F^I_{\mu\nu}\right> = q_I \left<F^{I,-}_{\mu\nu}\right> =\frac{i}{4} e^{\frac{K}{2}}q_I \overline{X}^{I}\left<W^-_{\mu\nu} \right> = \frac{1}{4} \overline{Z}(q) \left<W^-_{\mu\nu} \right>\;.
\label{qfstar1egw}
\ee 
By choosing the phase of the graviphoton background appropriately, specifically such that 
\be 
\mathrm{Im} \left(\; \overline{Z}(q) \; \left<W^-_{\mu\nu} \right> \;\right) = 0 \;,
\label{zwcalcond}
\ee 
we can realise the relation
\be 
q_I \left<F^I_{\mu\nu}\right> = \mathrm{Re} \left( q_I \left<F^{I,-}_{\mu\nu}\right> \right)\;.
\label{qfstar1egwfin}
\ee 
This, up to a factor of two, matches the relation (\ref{qfstar1}).\footnote{Note that if there are multiple states, with different phases for $Z(q)$, the background $\left<W^-_{\mu\nu} \right>$ can be chosen to cancel only one phase.}

%So in the $s \rightarrow 0$ limit, we can identify the graviphoton with ($i$ times) the $\left<\hat{F}^{0,-}\right>$ direction. 

%Now imagine that we integrate out in the Euclidean background (\ref{gyoncharothapb}), a unit charged state, but then try to interpret it as an integrating out calculation in the Lorentzian background (\ref{onshelcondL}). We can write this as
%\be
%\left. \hat{q}_I \left<\hat{F}^{I,-}_{\mu\nu} \right> \right|_{\mathrm{Euc}} = \left. q_I \left<F^{I,-}_{\mu\nu} \right> \right|_{\mathrm{Euc}} =\left.q_I \left<F^I_{\mu\nu} \right> \right|_{\mathrm{Euc}} 
%\rightarrow  \;\;\left. q_I \left<F^I_{\mu\nu} \right> \right|_{\mathrm{Lor}} \;.
%\ee
%So the charges $q_I$ appear complex because we are interpreting standard unit charges $\hat{q}_I$ in the (complex) Euclidean background, as complex charges in a real Lorentzian background. 

While the K-point state couples to the Euclidean background the same way as the conifold state, we are interested in interpreting the integrating out calculation in a Lorentzian action. If we consider how the state couples to the Lorentzian background, so the analogue of (\ref{concouback}), we have 
\be
\text{K-point, Lorentzian}\;:\;q_I \left<F^{I}_{\mu\nu}\right> 
=  \frac{i\sqrt{c}}{4\sqrt{-\kappa_0\log |s|}} \Big( \bar{s} \left< W^-_{\mu\nu} \right>  - \left( 2\; i \;\mathrm{Im\;}\tau + s\right)\left< W^+_{\mu\nu} \right>\Big) + ... \;.
\label{concoubackK}
\ee 
We see that the coupling to $\left< W^-_{\mu\nu} \right>$ is the same as the Euclidean background. In particular, it vanishes in the $s \rightarrow 0$ limit (like in the conifold). However, the complex charges mean that the coupling to $\left< W^-_{\mu\nu} \right>$ and to $\left< W^+_{\mu\nu} \right>$ are not conjugates. They are different. In particular, crucially, the coupling to $\left< W^+_{\mu\nu} \right>$ does not vanish in the $s \rightarrow 0$ limit. 

Usually, we would expect to obtain the central charge and its conjugate as the respective couplings to $\left< W^+_{\mu\nu} \right>$ and $\left< W^-_{\mu\nu} \right>$. Because here they are different we should assign to each one a {\it chiral central charge} $Z(q)_{\pm}$. From (\ref{concoubackK}) we read off
\be 
\text{K-point\;}:\;\;Z_- = -i e^{\frac{K}{2}} \sqrt{c} \;s   \;\;,\;\;  Z_+ =  -i e^{\frac{K}{2}} \sqrt{c} \left(\;2 \;i\; \mathrm{Im\;} \tau + s   \right) \;.
\label{chicencck}
\ee 
These charges are the holomorphic objects we would expect to arise as contributions to the prepotential from integrating out, and indeed we discuss this in section \ref{sec:intprewfgh}. 

The fact that $Z_+ \neq Z_-$ for the K-point state can be taken as its key characteristic. The complex-charges are really a reflection of this. To our knowledge, this is a completely new type of BPS state.   

What mass should we associate to such a state? This is not clear. We have used so far 
\be 
M^2 = \left|Z_-\right|^2 =  c \;e^{K} \left|s\right|^2 \;.
\label{massBPSstad}
\ee  
However, we could also consider the more natural symmetric mass
\be 
M_{\mathrm{Sym}}^2 = \left|\overline{Z}_+Z_-\right| = 2 \; \left| \mathrm{Im\;} \tau\right|c\; e^K \left|s\right| + ... \;.
\label{massBPSNstad}
\ee 
In practice, the difference between the two is not important at this stage of investigations since both vanish as a power of $|s|$ , but it would be very interesting to understand this better.

\subsection{The gauge kinetic matrix from integrating out}

As discussed at the start of this section, the only integrating out calculation we know how to do is a background field calculation in Euclidean space with complex gauge fields. Therefore, from the perspective of its charges under the gauge fields, the integrated out BPS state at the K-point carries charges (\ref{kpochabpeuc}) under $\hat{F}^{I,-}_{\mu\nu}$, so the dynoic charges (\ref{rotfhqfi}) under the $F^I_{\mu\nu}$ . 

It is important to understand that, for both the conifold and the K-point, the fact that the integrated out state induces a term in the effective action of the form $\left<W^{-}_{\mu\nu}\right>\left< W^{-,\mu\nu}\right>$, whose coefficient vanishes at leading order, is sufficient to fix the induced form, from integrating it out, of the full gauge kinetic matrix. Let us describe this.

The graviphoton background (\ref{onshelcondL}), or (\ref{onshelcondE}), is a projection from all the gauge field strengths onto a single direction. It would then seem that an integrating out calculation in such a background would not be sufficient to probe the full gauge coupling matrix, but only some particular combination of its entries. However, when combined with the information on the periods, so the $X^I$, the full matrix can be extracted. This follows from the coefficient in (\ref{n2lfacasdgp}) in front of $\left<W^{-}_{\mu\nu}\right>\left< W^{-,\mu\nu}\right>$, from which we can extract the combination $\mathcal{N}_{IJ} X^I X^J$ which is nothing but the prepotential
\be 
\mathcal{N}_{IJ} X^I X^J = 2F(X) \;.
\ee 
From the prepotential, we can take derivatives to reach the full $\mathcal{N}_{IJ}$. This is the special property of the graviphoton background, it captures all the information. 

From the integrating out perspective, we obtain the leading logarithmic part of the gauge kinetic matrix. Then, because the charge of the BPS state under the graviphoton background vanishes, we know that the contribution to the coefficient in front of $\left<W^{-}_{\mu\nu}\right>\left< W^{-,\mu\nu}\right>$ from integrating out the state is such that
\bea 
\text{Conifold\;}&:&\;\left(1,s \right) \cdot \mathrm{Im\;}\mathcal{N} \cdot \left( \begin{array}{c} 1 \\ s \end{array} \right)  \sim  s^2 \ln |s| \;\; \implies \;\; \mathrm{Im\;}\mathcal{N} \sim  \left( \begin{array}{cc} 0 & 0 \\ 0 & 1\end{array}\right) \ln |s|\;, \nn \\
\text{K-point\;}&:&\;\left(1,\tau \right) \cdot \mathrm{Im\;}\mathcal{N} \cdot \left( \begin{array}{c} 1 \\ \tau  \end{array} \right)  = 0 \;\; \implies \;\; \mathrm{Im\;}\mathcal{N} \sim \left( \begin{array}{cc} a & b \\ b & c \end{array} \right)\ln |s|\;.
\eea
So, the leading dependence on $s$ of the graviphoton background coefficients, supplemented by the symmetry properties of ${\cal N}_{IJ}$ , is sufficient to fix the contribution as proportional to the actual leading form of the gauge kinetic matrix. So the full matrix, and not only a single component of it. We therefore see that the leading form of the gauge kinetic matrix is reproduced from integrating out the BPS state. This is a complimentary, and in some sense more sophisticated, alternative to the analysis in section \ref{sec:intcomchss}. 

\subsection{The prepotential from integrating out}
\label{sec:intprewfgh}

While the coupling of the BPS state at the K-point to the graviphoton background can be made the same as the conifold state, its mass is somewhat different. Indeed, we saw in section \ref{sec:charparbackw} that the K-point state has chiral central charges $Z_{+} \neq Z_{-}$. Relatedly, the form of the prepotential for the K-point behaves at leading order as $s \log s$ (\ref{KPrepo}) rather than the $s^2 \log s$ behaviour of the conifold (\ref{ConifoldPrepo}). Indeed, we can write the logarithmic terms in the prepotential, the ones associated to the leading gauge kinetic term behaviour, as (see, (\ref{ConifoldPrepo}), (\ref{KPrepo}) and (\ref{Fsx33}))
\bea
\text{Conifold\;} \;&:&\;\; F(s) \supset -\frac{1}{4\pi i} k \; s^2 \log s\;, \nn \\
\text{K-point\;} \;&:&\;\; F(s) \supset -\frac{1}{4\pi i} c\;s\left(s + 2i\mathrm{Im\;}\tau \right) \log s\;.
\label{prepotntr}
\eea 
Taking the (equal) chiral central charges for the conifold state as
\be 
\text{Conifold\;}:\;\;Z_- = Z_+ = -i e^{\frac{K}{2}} s \;,
\ee 
we can write the prepotential terms (\ref{prepotntr}), at leading order, in an identical form
\be
F(s) \supset \frac{1}{8\pi i} \; N \; e^{-K} Z_+ Z_-   \log \left(e^{-K} Z_-^2 \right) \;.
\label{prepotntrsuf}
\ee
Here $N$ is the number of states that are integrated out, so $N=k$ for the conifold, and $N=1$ for the K-point.

We therefore see that the relation between the conifold and the K-point is exact in the sense that we can associate the prepotential term (\ref{prepotntrsuf}) to integrating out a state with chiral central charges $Z_{\pm}$. The difference between the two is then that at the K-point the state has $Z_+ \neq Z_-$.\footnote{Note that we could consider the more symmetric form for (\ref{prepotntrsuf}) : $\frac{1}{8\pi i} \; N \; e^{-K} Z_+ Z_-   \log \left(e^{-K} Z_- Z_+ \right)$. This would then require taking $N = 2$ for the K-point.}

The chiral central charges (\ref{chicencck}) imply the rather remarkable difference between the conifold and the K-point, which is that integrating out the state at the K-point yields an infinite distance singularity in the moduli space. This arises from the linear term $s \log s$ in the prepotential, coming from the cross term between the central charges. The relation between this term and the infinite distance is shown in section \ref{sec:mirinf}. 
%This difference has crucial implications for our understanding of the Swampland distance conjecture, as discussed in section \ref{sec:swdis}. 

\subsection{A geometric realisation of the graviphoton background}

As explained in section \ref{sec:ansfgbacks}, it is not possible to maintain simultaneously supersymmetry, flat spacetime, and real gauge field-strength backgrounds in a four-dimensional setting. However, in the case of only the graviphoton, so with no vector multiplets, there is a way to do so in five dimensions, so on $\mathbb{R}^{1,4}$ \cite{Gauntlett:2002nw,Dedushenko:2014nya}. More generally, since the background that is turned on is solely along the graviphoton direction, we should think generally of only a single condition of reality. So the general case is closely related to the case of the pure graviphoton. 

Understanding how to embed the theory in a higher dimensional one yields a geometric description of the physics of the graviphoton background, and leads to some insights into the microscopic physics behind it. This section is about introducing and developing this description.

The known supersymmetric background for an anti self-dual graviphoton is a Minkowski signature background of the form $\mathbb{R}^{1,4}$, found in \cite{Gauntlett:2002nw,Dedushenko:2014nya}. This background takes the form
\be 
ds^2 = - \left(dt - U_{\mu} dx^{\mu} \right)^2 + dx^{\mu}dx_{\mu} \;,
\label{godesol}
\ee 
where $U_{\mu} = \frac12 \left< T^-_{\nu\mu} \right> x^{\nu}$ with $\left< T^-_{\nu\mu} \right>$ a constant anti self-dual graviphoton background of the five-dimensional graviphoton.\footnote{It is anti self-dual with respect to the four Euclidean dimensions rather than the full $\mathbb{R}^{1,4}$.} The background (\ref{godesol}) is similar to the famous G\"odel solution \cite{RevModPhys.21.447}. We therefore refer to it as a G\"odel background. Of course, it differs in some ways from the G\"odel metric, for example, the spatial metric is flat. But it does share the crucial mixed term between time and space, and the closed timelike curves associated to this. It is therefore a rather exotic background: the imaginary gauge fields are exchanged for closed timelike curves. 

To make contact with the Euclidean four-dimensional supergravity background, we Wick rotate (\ref{godesol}) and compactify the time direction to reach $S^1 \times \mathbb{R}^4$. This yields a background
\be 
ds^2 = e^{2\sigma}\left(dy - \frac{i}{4} e^{-\frac{3\sigma}{2}} V_{\mu} dx^{\mu} \right)^2 + dx^{\mu}dx_{\mu} \;,
\label{eucsol} 
\ee  
where we Wick rotated and compactified time to a circle of radius $e^{\sigma}$ as
\be 
t = -i e^{\sigma} y \;,
\label{wicktimec}
\ee 
and 
\be 
V_{\mu} = 4 e^{\frac{\sigma}{2}} U_{\mu} = -\frac12 \left<W^{-}_{\mu\nu}\right> x^{\nu}\;,
\ee 
is the four-dimensional graviphoton background, with the constant field strength $\left<W^{-}_{\mu\nu}\right>$. In the four-dimensional setting, we see that if we demand to have a real spacetime metric, we need $\left<W^{-}_{\mu\nu}\right>$ to be imaginary \cite{Dedushenko:2014nya}. 

We can interpret this as a geometric manifestation of the condition (\ref{onshelcondE}), which for a pure graviphoton setting reads 
\be 
\mathrm{Euclidean\;\;:\;\;}\left<F^{0}_{\mu\nu}\right>  = \left<F^{0,-}_{\mu\nu}\right> =\frac{i}{4} e^{\frac{K}{2}} \left<W^-_{\mu\nu}\right> \;.
\label{onshelcond20lcs}
\ee 
We can take $\left<F^{0}_{\mu\nu}\right>$ as the real Kaluza-Klein gauge field background, coming from the (M-theory) circle direction. Then the condition (\ref{onshelcond20lcs}) is the matching condition between the standard (M-theory) circle reduction, and the supersymmetric graviphoton background of the G\"odel spacetime (\ref{godesol}). 

The identification (\ref{onshelcond20lcs}) is an M-theory uplift of the type IIB physics, through the type IIA mirror. That is, it is an identification of $F^{0,-}_{\mu\nu}$ with the M-theory circle in the type IIA uplift. This is indeed the correct identification in the large complex-structure limit. We can see this by looking at the asymptotic form of the graviphoton (\ref{gravWdef}). That is
\bea
\text{LCS\;} \;&:&\;\; \lim_{s \rightarrow 0}\left(-X^I G_{I,\mu\nu} + F_I F^I_{\mu\nu} \right) \sim \left(\log |s| \right)^3 \left(  F^{0,-}_{\mu\nu} - \frac{3 i}{\log |s|} F^{1,-}_{\mu\nu}\right)\;.
\label{plimiasymgodlcs}
\eea 
In this uplift, the integrated out states are such that the particle worldline picture is mapped to M2 branes wrapping the M-theory circle, while a dynamical field theory picture would correspond to the Poisson dual description with Kaluza-Klein modes along the M-theory circle, which are $D2-D0$ branes. This is a geometric realisation of the duality between integrating out $D2-D0$ bound states and worlsheet instantons of the fundamental string \cite{Gopakumar:1998jq,Dedushenko:2014nya,Hattab:2024ssg}. Using this uplift, we see that we could exchange the Euclidean imaginary graviphoton background calculation for a Lorentzian calculation with a real background but in a G\"odel-type spacetime. We do not know how to do either one very well, but at least we expect that at weak backgrounds $\left<W^-_{\mu\nu}\right> \rightarrow 0$ , the calculations should be correct \cite{Dedushenko:2014nya}. The results of \cite{Hattab:2024ewk,Hattab:2024yol,Hattab:2024ssg} suggest that the calculation should also hold for very large backgrounds $\left<W^-_{\mu\nu}\right> \rightarrow 
\infty$ , and in some cases for possibly all backgrounds. 

In the K-point region we do not have a simple dual supergravity picture into which the solutions (\ref{godesol}) and (\ref{eucsol}) could be embedded. However, we can gain some insights by studying the supergravity. The graviphoton alignment limit (\ref{plimiasymgodlcs}) becomes now
\bea
\text{K-point\;} \;&:&\;\; \lim_{s \rightarrow 0}\left(-X^I G_{I,\mu\nu} + F_I F^I_{\mu\nu} \right) \sim  \left( F^{0,-}_{\mu\nu} - \frac{1}{\tau} F^{1,-}_{\mu\nu} \right) \log |s|\;.
\label{plimiasymgodK}
\eea 
This suggests that a geometric interpretation is one where instead of a circle associated to $F^{0,-}_{\mu\nu}$ we have a torus, with a complex-structure $\tau'$ and holomorphic coordinate $w'$. So we can think of (\ref{plimiasymgodK}) as the holomorphic direction of a torus
\be 
dw' = dx_0 + \tau' \; dx_1 \;, \;\; \tau' = -\frac{1}{\tau} \;.
\ee 
After performing an $SL(2,Z)$ S-transform, sending $\tau' \rightarrow \tau$ and $w' \rightarrow w$, we have
\be 
dw = dx_1 - \tau \; dx_0 \;.
\ee 
This is very similar to the combination (\ref{rotfhqfi}). Indeed, identifying 
\be 
dx_0 \leftrightarrow F^{0,-}_{\mu\nu} \;\;,\;\; dx_1 \leftrightarrow F^{1,-}_{\mu\nu}\;,
\ee 
which corresponds to the Euclidean background, we have for the Lorentzian background (so after complexifying back the $F^{I,-}_{\mu\nu}$)
\bea 
\mathrm{Re} \;d\bar{w} &=& \mathrm{Re} \;dx_1 - \Big(\mathrm{Re} \;\tau\; \mathrm{Re} \;dx_0 + \mathrm{Im} \;\tau\; \mathrm{Im} \;dx_0\Big)\;, \nn \\
&\rightarrow & \frac12 \left( F^1_{\mu\nu} - \left[ \mathrm{Re\;}\tau \;F^{0}_{\mu\nu} + \mathrm{Im\;}\tau \;\tilde{F}^{0}_{\mu\nu}\ \right) \right] \;,
\eea 
which is precisely the combination appearing in (\ref{rotfhqfi}).

This suggests that the K-point region has physics similar to the large complex-structure region, in the sense that one could identify a circle direction with the graviphoton background. And charged states as particle worldlines on that circle. Possibly there is even a Wick rotation to a real background in some G\"odel-like spacetime. The difference would be analogous to instead of taking a single circle direction, we would take a special Lagrangian one-cycle in a torus with complex structure $\tau$. The calibration angle of the special Lagrangian would then be mapped to the phase of the graviphoton background. The special Lagrangian condition would then be mapped to the condition (\ref{zwcalcond}). 

There is a hint that this is indeed the correct microcopic physics from the M-theory to Heterotic duality \cite{Dedushenko:2014nya}
\be
\begin{array}{c}
	\frac{\rule[-0.8ex]{0pt}{2.5ex}\text{\normalsize M-Theory}}{\rule{0pt}{3ex}\text{\normalsize $\widetilde{Y}$}} \simeq \frac{\rule[-0.8ex]{0pt}{2.5ex}\text{\normalsize Heterotic}}{\rule{0pt}{2.8ex}\text{\normalsize $K3\times S^1$}} \;\;.
\end{array} 
\label{chaduM}
\ee
Here $\tilde{Y}$ denotes a Calabi-Yau that is both elliptic and K3 fibered. 
The duality (\ref{chaduM}) implies of course
\be
\begin{array}{c}
	\frac{\rule[-0.8ex]{0pt}{2.5ex}\text{\normalsize M-Theory}}{\rule{0pt}{3ex}\text{\normalsize $\widetilde{Y} \times S^1$}} \simeq \frac{\rule[-0.8ex]{0pt}{2.5ex}\text{\normalsize Heterotic}}{\rule{0pt}{2.8ex}\text{\normalsize $K3\times T^2$}} \;\;.
\end{array} 
\label{chaduM2}
\ee 
This duality is the natural one for the K-point, so replaces the M-theory type IIA circle duality. The $S^1$ on the M-theory side in (\ref{chaduM2}) should then be identified with a special Lagrangian cycle in the $T^2$ on the Heterotic side, which is naturally assigned a complex-structure $\tau$. Note that in the duality (\ref{chaduM2}) the $S^1$ is geometric on the Heterotic side, as opposed to the string coupling in the duality with IIA at large complex-structure. This means that the M-theory dual is not expected to be in the supergravity regime.

We hope to return to understanding the microscopic physics associated to the perspective on the K-point region that we have developed here in future work.

\section{Distances in moduli space and Swampland conjectures}
\label{sec:swdis} 

In this work we presented an analysis of the physics around the K-point in complex-structure moduli space. We found evidence that the infrared theory has a light BPS state integrated out such that the leading behaviour of the gauge kinetic matrix, and the leading dependence of the prepotential on the modulus $s$, are induced by integrating the state out. That is, they should be understood as threshold correction type physics. 

This is a new perspective on the K-point, even though its physics was analysed already in \cite{Grimm:2018ohb}, and in earlier works. The picture, prior to this work, is of a theory that behaves much like the large complex-structure region in moduli space, and not like the conifold region. This was also the picture presented in the more recent analysis \cite{Friedrich:2025gvs}, where a weakly-coupled Heterotic dual was proposed. The new perspective we are proposing is much more radical, and relies on ideas that we do not yet fully understand. The most striking being that the would-be BPS state that is integrated out would have apparently complex charges. We have presented evidence for these ideas, and in this section proceed to study what the implications would be if these ideas are correct. More specifically, we are interested here in the implications for the Distance Conjecture \cite{Ooguri:2006in} as part of the Swampland program \cite{Vafa:2005ui,Palti:2019pca}.

\subsection{An emergent infinite distance}
\label{sec:mirinf}

Integrating out the BPS state in the conifold region yields the leading (logarithmically divergent in $s$) behaviour of the gauge kinetic matrix (\ref{congkf}). However, it only has a minor effect on the moduli space metric, and in particular, leaves the conifold locus at finite distance in moduli space. By contrast, the K-point is at infinite distance in moduli space. The natural question is therefore whether this divergence is associated to integrating out the BPS state or not. In this section we show that it is.

Before technically showing this, there is a simple conceptual way to understand why that must be so. Points of infinite distance in the moduli space are determined by the behaviour of the monodromy matrix ${\bold T}$ in (\ref{tdef}) \cite{Grimm:2018ohb}. This monodromy is associated to the leading behaviour of the gauge kinetic matrix. As explained in sections \ref{sec:BPSint} and \ref{sec:elecmag}, integrating out the state yields the leading gauge kinetic matrix behaviour. It therefore determines the monodromy, and by association, should imply the infinite distance nature of the moduli space singularity. 

To see the infinite distance explicitly, note that the infinite distance behaviour of K-points can be tracked to the divergence of the Kahler potential in (\ref{Kkpoi}). The supergravity kinetic terms (\ref{supkint}) are given by 
\be 
- g_{z\bar{z}} \left|\partial z \right|^2  = - \left(\partial_z \bar{\partial}_z K \right) \left|\partial z \right|^2 \;.
\ee 
Because $z$ and $s$ only differ by a constant (\ref{oneparasing}), we can write the kinetic terms as
\be 
-\lim_{s \rightarrow 0} \; g_{z\bar{z}} \left|\partial z \right|^2 =  -\lim_{s \rightarrow 0} \; \left(\partial_s \bar{\partial}_s K \right) \left|\partial s \right|^2 = - \frac{1}{4|s|^2 \left(\log |s| \right)^2} \left|\partial s \right|^2 \;.
\label{kinztos}
\ee 
If we decompose $s$ into a radial mode and a phase
\be 
s = |s|e^{i \theta} \;,
\label{smodph}
\ee 
then the kinetic terms (\ref{kinztos}) are
\be 
- \frac{1}{4 |s|^2 \left(\log |s|\right)^2} \Big( \left(\partial |s|\right)^2 + |s|^2 \left(\partial \theta \right)^2 \Big) = -\frac14 \Big( \left(\partial \log \left(-\log |s| \right)\right)^2 + \frac{1}{\left(\log |s|\right)^2}\left(\partial \theta \right)^2 \Big) \;.
\ee 
We therefore see that the canonically normlized scalar field with which to measure the distance is 
\be 
\Delta = \frac{1}{\sqrt{2}} \log \left( -\log |s| \right)  \;.
\label{deltakp}
\ee 
Therefore the K-point limit $|s|\rightarrow 0$, is at infinite distance $\Delta \rightarrow \infty$. 

We can track the infinite distance to the $\log |s|$ term in (\ref{Kkpoi}), and then to the $s \log s$ term in the prepotential (\ref{KPrepo}). This $s \log s$ term arises from integrating out the BPS state, so as a threshold correction to the prepotential (\ref{prepotntrsuf}). Therefore, integrating out the state is what induces the infinite distance. If we were to remove that the contribution (\ref{prepotntrsuf}), then there would be no divergence in the distance. 
In this sense the infinite distance is emergent, as proposed in this context in \cite{Grimm:2018ohb,Palti:2019pca}.\footnote{The more general Emergence Proposal \cite{Palti:2019pca} is based on ideas in \cite{Harlow:2015lma,Heidenreich:2017sim,Grimm:2018ohb,Heidenreich:2018kpg} and we refer to \cite{Palti:2020tsy,Marchesano:2022avb,Castellano:2022bvr,vandeHeisteeg:2022btw,Cribiori:2022nke,Marchesano:2022axe,Blumenhagen:2023tev,Castellano:2023qhp,Cribiori:2023ffn,Blumenhagen:2023yws,Blumenhagen:2023xmk,Seo:2023xsb,Calderon-Infante:2023ler,Calderon-Infante:2023uhz, Castellano:2023aum,Castellano:2023jjt,Castellano:2023stg,Cota:2023uir,Hattab:2023moj,Casas:2024ttx,Blumenhagen:2024ydy,Hattab:2024thi,Blumenhagen:2024lmo,Hattab:2024ewk,Artime:2025egu,Hattab:2024ssg,Hattab:2024chf} for a selection of works studying these ideas in various settings in string theory.} 

The scale below which the distance is emergent is set by the mass of the BPS state. 
In terms of this distance the asymptotic behaviour of the mass $M(q,s)$ is \footnote{We take the mass of the state as in (\ref{massBPSstad}), rather than (\ref{massBPSNstad}). Qualitatively, there is no difference for this discussion.}
\be 
\lim_{s \rightarrow 0} M(q,s) = \lim_{s \rightarrow 0} |Z\left(q,s\right)| = \frac{\sqrt{c}}{\sqrt{-\kappa_0 \log |s|\;}}\; |s| = \sqrt{\frac{c}{\kappa_0}} e^{-\left( e^{\sqrt{2} \Delta} +\frac{1}{\sqrt{2}}\Delta\right)} \;.
\label{sdedd}
\ee 
So it is doubly exponentially light in the distance $\Delta$. The scale becomes light extremely fast when approaching the K-point, exponentially faster than any integer charged states, which become light only exponentially in the distance.

\subsection{Implications for the Distance Conjecture}
\label{sec:scalandgen}

To understand the role that the K-point plays with respect to the Distance Conjecture \cite{Ooguri:2006in}, it is informative to first examine the role that the conifold plays with respect to the Magnetic Weak Gravity Conjecture (WGC) \cite{Arkani-Hamed:2006emk}.

\subsubsection{The conifold and the Magnetic WGC}

The (magnetic) WGC proposes that the gauge coupling $g$ of a $U(1)$ gauge symmetry in the effective theory sets a cutoff scale of the theory. For the sake of this discussion, we formulate this statement by an explicit choice of what occurs at this cutoff as follows.\footnote{We refer to \cite{Palti:2019pca,Palti:2020mwc,vanBeest:2021lhn,Harlow:2022ich} for reviews of the Weak Gravity Conjecture and discussions of various formulations of both electric and magnetic versions of it.} There exists an infinite tower of charged particles, with characteristic mass scale $m_{\infty}$ , such that 
\be 
m_{\infty} \sim g\; M_p \;.
\label{wgcor}
\ee 
Now consider approaching the conifold locus in moduli space. In that case, we have that the gauge coupling (in the infrared) behaves schematically as (\ref{congkf})
\be 
\frac{1}{g^2} \sim -\frac{1}{\alpha} \log |s| \;,
\ee 
where $\alpha$ is some positive constant.
Therefore, (\ref{wgcor}) predicts a tower of charged particles with mass
\be 
m_{\infty} \sim g\;M_P \sim \frac{1}{\sqrt{-\frac{1}{\alpha}\log |s|}} \;M_p \;.
\label{mwgcconn2}
\ee 
This does not hold.

There are two reasons why (\ref{mwgcconn2}) is not true, which shed light on how we should think of formulating the WGC. The first is that the effective supergravity description breaks down at a scale much lower than (\ref{mwgcconn2}). The conifold limit corresponds to a three-cycle shrinking to zero size, but to have a supergravity description we need to keep the volume of the three-cycle large in string units (and the string coupling weak). This requires a double scaling limit where the CY volume is scaled when approaching the conifold locus as  
\be 
\left(\mathrm{Vol}_{\mathrm{CY}}\right)^{\frac12} \sim |s|^{-1} \sim e^{\frac{\alpha}{g^2}}\;.
\ee  
We therefore have a cutoff on the theory given by the Kaluza-Klein scale of the Calabi-Yau of 
\be 
\Lambda_{\mathrm{KK}} \sim \left(\mathrm{Vol}_{\mathrm{CY}}\right)^{-\frac16} M_s \sim \left(\mathrm{Vol}_{\mathrm{CY}}\right)^{-\frac23} M_p \sim e^{-\frac{4\alpha}{3g^2}} M_p \;,
\label{brawgcon}
\ee 
where $M_s$ is the string scale. Because $\Lambda_{\mathrm{KK}}$ is exponentially lower than $m_{\infty}$ in (\ref{mwgcconn2}), we cannot use the effective value of $g$ in the infrared to determine physics at the scale $m_{\infty}$. The theory completely changes before that scale. 

There is a second reason why (\ref{mwgcconn2}) does not hold, which is more general, and so more directly generalized to the Distance Conjecture and the K-point. The infrared gauge coupling is dominated by the threshold correction from the integrated out BPS state. This means that it is dominated by the running from the ultraviolet to the mass of the BPS state $M$, which behaves as
\be 
M \sim |s| M_p \sim e^{-\frac{\alpha}{g^2}} M_p \;.
\label{conmassbsw}
\ee 
If we consider the value of the gauge coupling at the scale $m_{\infty}$ in (\ref{mwgcconn2}), it would be completely different because of the strong running between the scale $m_{\infty}$ and the scale $M$.   
 
 The conifold example therefore teaches us about how the Magnetic WGC should be formulated, at least so that there is not such a simple counter-example. One way to formulate it is to say that in (\ref{wgcor}) the value of the gauge coupling should be the one at the associated scale, that is at $m_{\infty}$. This is how it is usually thought of. However, it is important to understand that such a formulation is not predictive in any sense, that is, we are required to know the physics all the way up to the predicted scale of new physics before we can predict what the scale of the new physics is. In other words, if we know the theory above the scale (\ref{wgcor}), and know that there is no tower of particles, then we can say that the theory is in the Swampland. But if we do not know the theory at that scale, then we cannot predict that there will be a tower of particles there. 
 
 Having to know the theory to the scale $m_{\infty}$ is particularly problematic since that scale is associated to an infinite tower of states. This is because it is known that there is no consistent cutoff possible in the middle of such a tower. Therefore, any effective theory which describes the physics up to the tower scale, must also hold to the full ultraviolet, or at least until some new duality frame can be used, in which case the original duality frame used to formulate the theory breaks down. So, for example, we can only write an effective description at a Kaluza-Klein scale if we start from a higher dimensional theory. 
 
 The conifold example teaches us that the WGC should be formulated only as a statement about scales that we already know the physics for. Therefore, it cannot be formulated as a statement about an infinite tower of states and also be applied to an effective theory. In this sense, as a statement on effective theories, it should be formulated as an upper bound on the cutoff of the effective theory $\Lambda$. So as the statement
 \be 
 \Lambda < g\left[\Lambda\right] M_p \;,
 \label{Lmwgcfm}
 \ee  
 where $g\left[\Lambda\right]$ denotes evaluating the gauge coupling at the scale $\Lambda$. Further, importantly, the upper bound in (\ref{Lmwgcfm}) cannot be claimed to be approximately saturated. This is obvious for general $\Lambda$. But we could imagine a setup where we take $\Lambda$ as the maximum cutoff possible within an effective theory, so the maximum scale for which we can write an effective theory within a given duality frame. And then claim that this scale is such that (\ref{Lmwgcfm}) is approximately saturated. However, the conifold example rules out such a possibility, there the breakdown of the effective description is exponentially lower (in terms of $g$) than the scale (\ref{Lmwgcfm}).
 
 We therefore see that the conifold example shows that the formulation (\ref{Lmwgcfm}) is the relevant statement for applying to a consistent effective field theory. 
 
 \subsubsection{The K-point and the Distance Conjecture}
 
 We now turn to the Distance Conjecture \cite{Ooguri:2006in}. The Distance Conjecture is formulated in a similar way to (\ref{wgcor}). It proposes that there exists an infinite tower of states, with mass scale $m_{\infty}$, which become exponentially light as a function of the moduli space distance $\Delta$ when approaching infinite distances
\be 
m_{\infty} \sim e^{-\gamma\;\Delta} \;M_p \;,
\label{disconjor}
\ee  
where $\gamma$ is some order one parameter. 
 
In general, all of the discussion above of the Magnetic Weak Gravity Conjecture applies equally to the Distance Conjecture, where instead of the gauge coupling $g$ we consider the moduli space distance $\Delta$. Therefore, we expect that the only formulation of the Distance Conjecture that can be applied to an effective field theory is of the form analogous to (\ref{Lmwgcfm}), that is
\be 
 \Lambda < e^{-\gamma\;\Delta\left[\Lambda\right]} M_p \;,
 \label{Lmdcfm}
 \ee  
 where $\Delta\left[\Lambda\right]$ refers to the distance as calculated at the scale $\Lambda$. Again, as in (\ref{Lmwgcfm}), the inequality sign could be strongly violated.\footnote{In fact, this is how the Refined Distance Conjecture was formulated in \cite{Klaewer:2016kiy,Palti:2019pca}, explicitly stating that it allows for strong violations of the exponential behaviour.}
 %In generality, without further assumptions, there seems no consistent way to formulate a statement relating the distance to a tower of states, while allowing the statement to be applied in an effective field theory. 
 
We can consider the much stronger form 
\be 
m_{\infty} \sim e^{-\gamma\;\Delta\left[0\right]} \;M_p \;,
\label{disconjor0}
\ee
so with the distance calculated in the (infinite) infrared.
In practice, this is how the Distance Conjecture is used in the literature, and how it was formulated in \cite{Ooguri:2006in}. In generality, following the logic of the conifold analysis of the Weak Gravity Conjecture above, we may expect that (\ref{disconjor0}) cannot hold true. But, prior to this work, there was no candidate for a conifold-like example for the Distance Conjecture, since the conifold is at finite distance in moduli space while the Distance Conjecture refers to infinite distances. 

Note that, of course, one can consider (\ref{disconjor0}) where the mass scale $m_{\infty}$ is attributed to the energy of any state. In fact, such a version of the Weak Gravity Conjecture holds also for the conifold. There is an infinite tower of BPS states with increasing energy in the conifold limit starting from the scale $M$ in (\ref{conmassbsw}). They are just multi-particle states of the conifold particle. But the point is that $m_{\infty}$ in (\ref{disconjor0}) ( and in (\ref{wgcor}) ) is supposed to apply to independent (so weakly-coupled single-particle) fundamental states of the theory. This is what makes the discussion non-trivial. Indeed, the presence of BPS states of energies satisfying (\ref{disconjor0}) was proven in full generality in \cite{Grimm:2018ohb}, for any locus in the moduli space. But this should be taken as proving (\ref{Lmdcfm}), rather than necessarily (\ref{disconjor0}).

The picture of the K-point in moduli space that we have considered in this work plays the role for the Distance Conjecture that the conifold does for the (magnetic) Weak Gravity Conjecture. That is, if the picture of the K-point as being emergent from a BPS state is correct, then the distance in the infrared is dominated by the running down to the scale (\ref{sdedd}), so (analogously to (\ref{conmassbsw})) down to
\be 
M \sim e^{e^{-\beta\Delta\left[0\right]}} M_p \;,
\label{mkpobpsmassw}
\ee 
with $\beta$ some positive constant. Therefore, it is a counter example to the version (\ref{disconjor0}). In other words, the distance in the infrared is dominated by the threshold correction from the light BPS state, just as in the conifold, and this is attributed to strong running from the ultraviolet to the infrared.

A more robust way to say this is that, as in the Weak Gravity Conjecture and the conifold, the effective theory breaks down at the scale $M$ in (\ref{mkpobpsmassw}), rather than $m_{\infty}$ in (\ref{disconjor0}). There is therefore no sense in which we can treat the states at $m_{\infty}$ as fundamental.\footnote{BPS states of those energies do exist around the K-point region, but as solitonic configurations of the infrared effective theory.} It is a more robust statement because it makes no assumption about what the theory would look like above the scale $M$ in (\ref{mkpobpsmassw}). Nonetheless, we expect that any notion of a distance above that scale would be such that the distance to the K-point is finite in the ultraviolet. Note that, because the theory breaks down not by the appearance of an infinite tower of weakly-coupled fundamental states, a version of (\ref{disconjor0}) where we allow for a strong inequality also does not hold. The only version which holds is (\ref{Lmdcfm}).  

 This discussion shows that understanding the scenario of the K-point region that we have studied in this paper, which admittedly currently leaves many open questions, could be central to the Distance Conjecture, and the Swampland program in general. We therefore emphasise that this is the content of this analysis in the context of the Swampland program: we do yet claim that the K-point emergent infinite distance scenario we have developed is completely proven, but wish to emphasise that if it is indeed correct, then it has many important implications. 

\section{Discussion}
\label{sec:disc}

In this paper we studied the K-point region in one-parameter complex-structure moduli spaces arising from compactifications of type IIB string theory on Calabi-Yau manifolds. There is an effective four-dimensional supergravity description of this region, but it is one where all massive charged BPS states have been integrated out, and so is valid only in the (infinite) infrared. We analysed this supergravity description for clues as to how the finite energy scale theory should behave. We found a number of qualitative and quantitative similarities between the supergravity description of the K-point and that of the conifold locus in moduli space. 

The conifold locus is known to arise in the infrared from integrating out a BPS state, more precisely, the leading behaviour of the gauge kinetic terms and of the prepotential can be argued to arise from a single light BPS state. The similarities between the K-point and the conifold therefore suggest that the K-point also has an interpretation as arising from integrating out a light BPS state. 

We developed this picture of the K-point as arising from integrating out a BPS state. We found that the required state has certain exotic properties: it has apparently complex charges and it becomes light doubly-exponentially fast in the distance to the K-point. The complex charges could be rephrased as having different coupling to the self-dual and the anti self-dual components of the graviphoton. Indeed, from the perspective of integrating out the state in a supersymmetric background field computation, the different couplings are the defining feature of the state. While taking these properties reproduces the effective supergravity theory precisely, their meaning at the microscopic level remains to be better understood.

We also discussed the implications of this perspective on the K-point for the Distance Conjecture and the Swampland program. We argued that it is an example of an emergent infinite distance. It also amounts to a counter-example to the standard use of the Distance Conjecture. More precisely, it implies that the tower of states whose mass is exponentially decreasing cannot be understood as weakly-coupled fundamental states in any duality frame. In fact, this perspective on the K-point suggests that in the ultraviolet any notion of distance would be one where the K-point is at finite distance, rather than the infinite distance of the infrared theory.  We discussed a milder version of the conjecture which is compatible with this picture of the K-point.

We can also comment on whether we expect more such states in general. 
If the moduli space is not one-parameter, but multi-parameter, then for any infinite distance singularities there can also be other type F field theory states but with integral charges. These are naturally associated to various strongly-coupled sectors that could exist on the infinite distance singularity in moduli space. Such strongly coupled sectors are for example part of the picture in \cite{Monnee:2025msf}, and played an important role in the studies in \cite{Marchesano:2023thx,Marchesano:2024tod,Castellano:2024gwi,Blanco:2025qom}. The natural expectation is that integrating out these states does not generate the full infinite distance but only some part of the singularity. We expect that type F states which are associated to the infinite distance part of the singularity, or more directly, to the appropriate part of the monodromy matrix, require complex charges.

 In summary, we find that the K-point in moduli space behaves like it is associated with very rich physics, that can have much to teach us in general, and specifically in the context of the Swampland and of Emergence. It appears to be an exotic version of the conifold. We hope that further investigations will uncover the microscopic physics behind the fascinating behaviour of the supergravity in the infrared. We also expect that the physics is not unique to the K-point, but likely holds for any type II (and possibly type III) loci in moduli space (in the language of \cite{Grimm:2018cpv}). 
 
\vspace{10pt}
{\bf Acknowledgements}
%\vspace{10px}
\noindent
We would like to thank Ofer Aharony, Damian van de Heisteeg, Dieter L\"ust, Fernando Marchesano, Carmine Montella, Nicolo Petri, Timo Weigand, Max Wiesner and Matteo Zatti for very useful discussions. 
This work was supported by the German Research Foundation through a German-Israeli Project Cooperation (DIP) grant ``Holography and the Swampland" and by the Israel planning and budgeting committee grant for supporting theoretical high energy physics.

\appendix

\section{An example one-parameter Calabi-Yau: $\mathbb{P}^5[3,3]$}
\label{sec:examcyx33}

\begin{figure}
\centering
 \includegraphics[width=1.0\textwidth]{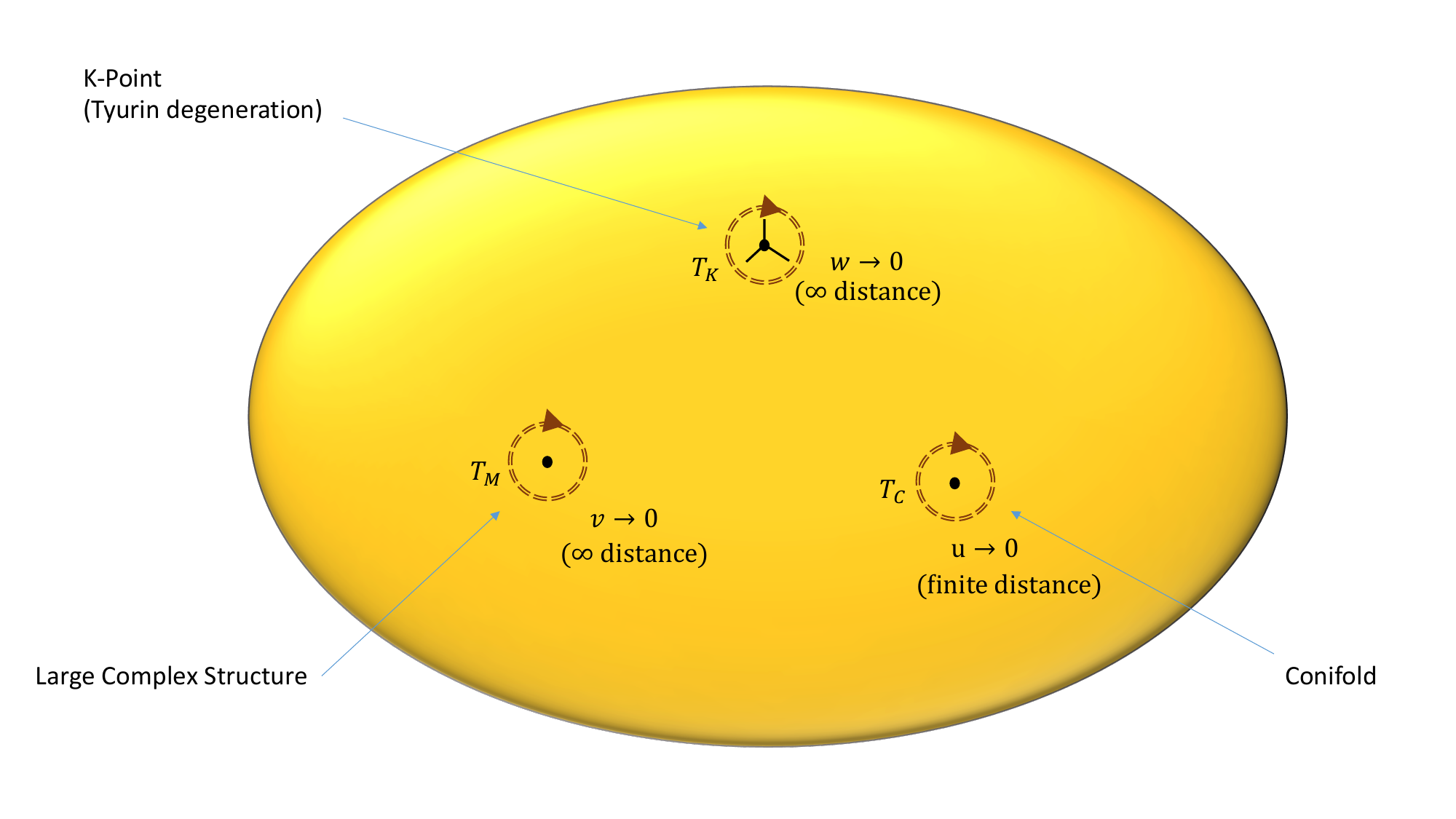}
\caption{Figure showing the moduli space of the $X_{3,3}$ Calabi-Yau, which has three singular loci. Each singular locus has a monodromy associated to it which acts on the period matrix upon circling the singular point.}
\label{fig:modulispace}
\end{figure}

In this section, we calculate the expressions of the period vector, prepotential, gauge kinetic matrix, Kahler potential and moduli space metric for the particular example of the one-parameter bi-cubic Calabi-Yau $\mathbb{P}^5[3,3]$, which is sometimes denoted as $X_{3,3}$. We refer to, for example, \cite{Joshi:2019nzi,Palti:2021ubp} for more details on the manifold. It is defined in terms of homogeneous coordinates $x_i$ in $\mathbb{P}^5$ and a complex structure moduli space coordinate $\psi$ by
\begin{align}
    \frac{x_1^3}{6}+\frac{x_2^3}{6}+\frac{x_3^3}{6}+\psi\; x_1x_2x_3 = 0\;,\\
    \frac{x_4^3}{6}+\frac{x_5^3}{6}+\frac{x_6^3}{6}+\psi \;x_4x_5x_6 = 0\;.
\end{align}
It is useful to introduce coordinates on the moduli space 
\be
w \equiv \psi^2 \;,\;\; v \equiv \frac{1}{\mu \psi^6} \;,\;\; u \equiv 1-v \;,
\label{codef}
\ee
where $\mu = 3^6$. The moduli space has three special points
\begin{itemize}
\item Large complex structure : $v = 0$,
\item K-point (Tyurin degeneration) : $w = 0$,
\item Conifold point: $u = 0$.
\end{itemize}
The first two of these are at infinite distance, while the conifold is at finite distance. The K-point locus $w=0$ has a $\mathbb{Z}_3$ orbifold symmetry which can be seen by noting that sending $\psi \rightarrow \psi \;e^{\frac{2 \pi i}{3}}$ can be undone by rotating the coordinates $x_i$ appropriately. The moduli space is illustrated in figure \ref{fig:modulispace}. 

The period vector can be written in the Frobenius basis in terms of Meijer G-functions \cite{Joshi:2019nzi}
\begin{equation}
\label{frob}
\tilde{\Pi} = \left(\begin{array}{cccc}
G_{4,4}^{1,4}\left(-\psi^6\;|
\begin{array}{c}
 1,1,1,1 \\
 \frac{1}{3},\frac{1}{3},\frac{2}{3},\frac{2}{3} \\
\end{array}\right)\\
G_{4,4}^{1,4}\left(-\psi^6\;|
\begin{array}{c}
 1,1,1,1 \\
 \frac{2}{3},\frac{1}{3},\frac{1}{3},\frac{2}{3} \\
\end{array}\right)\\
   G_{4,4}^{2,4}\left(\psi^6\;|
\begin{array}{c}
 1,1,1,1 \\
 \frac{1}{3},\frac{1}{3},\frac{2}{3},\frac{2}{3} \\
\end{array}
\right)\\G_{4,4}^{2,4}\left(\psi^6\;|
\begin{array}{c}
 1,1,1,1 \\
 \frac{2}{3},\frac{2}{3},\frac{1}{3},\frac{1}{3} \\
\end{array}
\right)
\end{array}\right)\;.
\end{equation}
The monodromy matrix in the Frobenius basis is defined at the K-point ($\psi =0$) by
\begin{equation}
    \tilde{\Pi}(e^{2\pi i }\psi^2) = \left(\mathbb{1}+\tilde{N}\right)\cdot\tilde{\Pi}(\psi^2)\;,
\end{equation}
where $\tilde{N}$ can be read from the expansion of (\ref{frob}) at $\psi = 0$ (see \cite{Palti:2021ubp}):
\begin{equation}
    \tilde{N} = -6\pi i \left(\begin{array}{cccc}
 0 & 0 & 0 & 0\\
 0 & 0 & 0 & 0\\
 (-1)^{-\frac13} & 0 & 0 & 0\\
 0 & (-1)^{-\frac23} & 0 & 0
\end{array}\right)\;.
\label{frobperiodmon}
\end{equation}
The Frobenius basis can then be mapped to the canonical integral symplectic basis (the mirror map basis) by the transformation
\begin{equation}
    \Pi' = T_{LF}\cdot\tilde{\Pi}\;,
\end{equation}
with $T_{LF}$ given by \cite{Joshi:2019nzi}
$$T_{LF}=
\left(
\begin{array}{cccc}
 -\frac{3 \left(5+i \sqrt{3}\right)}{8 \pi ^2} & \frac{3 \left(5-i \sqrt{3}\right)}{8 \pi ^2} & \frac{3 \left(\sqrt{3}+3 i\right)}{16 \pi ^3} & -\frac{3
   \left(\sqrt{3}-3 i\right)}{16 \pi ^3} \\
 \frac{-3-i \sqrt{3}}{4 \pi ^2} & \frac{3-i \sqrt{3}}{4 \pi ^2} & \frac{3 i}{8 \pi ^3} & \frac{3 i}{8 \pi ^3} \\
 -\frac{9 \left(3+i \sqrt{3}\right)}{8 \pi ^2} & \frac{9 \left(3-i \sqrt{3}\right)}{8 \pi ^2} & \frac{9 i}{8 \pi ^3} & \frac{9 i}{8 \pi ^3} \\
 \frac{-3-i \sqrt{3}}{8 \pi ^2} & \frac{3-i \sqrt{3}}{8 \pi ^2} & \frac{21 \left(1-i \sqrt{3}\right)}{8 \left(\sqrt{3}-5 i\right) \pi ^3} & \frac{3 i
   \left(4 \sqrt{3}+i\right)}{4 \left(\sqrt{3}-5 i\right) \pi ^3} \\
\end{array}
\right)\;.$$

\subsubsection*{Large complex-structure}

Expanded around the large complex structure point $z = (3\psi)^{-6} = 0$ , the period vector takes the following form 
\begin{equation}
   \Pi' \equiv   \left(
\begin{array}{c}
 X^{0'} \\
  X^{1'} \\
 F_0'\\
 F_1'\\
\end{array}
\right)=\left(
\begin{array}{c}
 1 \\
 \frac{ \log (v)}{2 \pi i} \\
 \frac{3  \log ^3(v)}{2(2 \pi i) ^3}+\frac{9  \log (v)}{4 (2\pi i)}-\frac{144  \zeta (3)}{(2\pi i)^3} +\mathcal{O}(v)\\
 -\frac{9 \log ^2(v)}{2(2 \pi i)^2}+\frac{ \log (v)}{2 (2\pi i) }+\frac{9}{4} +\mathcal{O}(v)\\
\end{array}
\right)\;.
\end{equation}
The symplectic pairing associated to this form, so as to match the supergravity formulae, is 
\begin{equation}
    \Omega = \left(\begin{array}{cccc}
 0& 0& -1& 0\\
 0& 0& 0& -1\\
 1& 0& 0& 0\\
 0& 1& 0& 0
\end{array}\right)\;.
\label{omegadefsy}
\end{equation}
So, for example, we have that 
\be 
K =  - \log \left[\frac{i}{4}\left(\Pi'\right)^T \cdot \Omega  
\cdot \overline{\Pi}'\right] \;.
\ee 
 
The monodromy matrix in the integral basis, $N'$, is then related to the Frobenius one by the transformation
\begin{equation}
    N' = T_{LF}\cdot \tilde{N}\cdot T_{LF}^{-1}  =\left(
\begin{array}{cccc}
 -9 & -6 & 6 & 3 \\
 -6 & -1 & 3 & 2 \\
 -18 & -3 & 9 & 6 \\
 -3 & -14 & 6 & 1 \\
\end{array}
\right)\;.
\end{equation}

\subsubsection*{The K-point}

To go to the appropriate basis for the K-point, we apply a further transformation by $ T_{KL}$ ,
\begin{equation}
    T_{KL} = -\Omega \cdot \left(
\begin{array}{cccc}
 12 & 7 & -8 & -4 \\
 1 & -8 & 2 & 0 \\
 3 & 2 & -2 & -1 \\
 0 & -3 & 1 & 0 \\
\end{array}
\right)\;.
\label{tklexpre}
\end{equation}
This is a symplectic transformation, so $ T_{KL}\in \text{Sp}(4,\mathbb{Z})$, under the symplectic pairing (\ref{omegadefsy}). This symplectic transformation takes the period vector to a basis 
\be 
\Pi = \left(
\begin{array}{c}
 \Pi^{0} \\
  \Pi^{1} \\
 \Pi^3\\
 \Pi^4\\
\end{array}
\right) = \Pi^0 \left(
\begin{array}{c}
 X^{0} \\
  X^{1} \\
 F_0\\
 F_1\\
\end{array}
\right)  = T_{KL}\cdot \Pi' \; , 
\label{pixif}
\ee 
where the K-point monodromy matrix $\bold{N}$ now becomes
\begin{equation}
    \bold{N} = -\left(\begin{array}{cccc}
 0& 0& 0& 0\\
 0& 0& 0& 0\\
 2& 1& 0& 0\\
 1& 2& 0& 0
\end{array}\right)\;.
\end{equation}
Since the $\Pi^I$ in (\ref{pixif}) use a different homogeneous gauge than the $X^I$ in the main sections $(X^0 = 1)$, we factorise the period vector by $\Pi^0$.

 In this basis, the period vector has the limiting behavior
\begin{equation}
    \lim_{\psi\rightarrow 0}\frac{\Pi^1}{\Pi^0} =  \lim_{\psi\rightarrow 0}\frac{X^1}{X^0} = \tau = e^{2\pi i /3}\;,
\end{equation}
meaning that we can solve for the prepotential $2F(X) =  X^0F_0+X^1F_1$ as an expansion in the parameter 
\begin{equation}
    s = \frac{X^1}{X^0}-\tau\;,
\end{equation}
which yields 
\begin{align}
\nonumber    F(s)\equiv F(X) =& -\frac{3}{2}+\frac{2 i}{\sqrt{3}}+\frac{4}{3}s- s\frac{\sqrt{3}}{2\pi}\left(\log\left[\frac{i}{\sqrt{3}}\left(\frac{\Gamma\left(\frac13\right)^2}{2\pi}\right)^6 s\right]-1\right)\\
    & -\frac{s^2}{2\pi i}\left(\log\left[\frac{i}{\sqrt{3}}\left(\frac{\Gamma\left(\frac13\right)^2}{2\pi}\right)^6 s\right]-2\right)+\mathcal{O}(s^3)\;,  
    \label{Fsx33}
\end{align}
where the higher-order corrections are devoid of any further logarithmic contributions. The Period vector then takes the following form
\begin{align}
\label{expanprepo}
    &\Pi/\Pi^0 = \left(\begin{array}{cccc}
 1\\
 \tau+s\\
 2F(s)-(s+\tau)\partial_s F(s)\\
 \partial_s F(s)
\end{array}\right)\; \\
\nonumber&\underset{s\rightarrow 0}{\sim} \left(\begin{array}{cccc}
 1\\
 \tau+s\\
 \frac{ 2 i \sqrt{3}-7}{3}+\frac{16\pi+3\sqrt{3}-9i}{12 \pi }s+\frac{s^2}{2\pi i }+\frac{2 i s-\sqrt{3}+3 i}{4 \pi }\log\left[\frac{i}{\sqrt{3}}\left(\frac{\Gamma\left(\frac13\right)^2}{2\pi}\right)^6 s\right]\\
 \frac{4}{3}+\frac{3}{2\pi i}s-\frac{\sqrt{3}-2is}{2\pi}\log\left[\frac{i}{\sqrt{3}}\left(\frac{\Gamma\left(\frac13\right)^2}{2\pi}\right)^6 s\right]
\end{array}\right)\;.
\end{align}

The gauge kinetic matrix can be calculated from the prepotential, and as an expansion in $s$ reads 
\begin{equation}
    \mathcal{N}_{IJ} = -\frac{\log |s|}{2\pi  i}\;\bold{B}_{IJ}+\mathcal{O}(1)\;,
\end{equation}
with 
\be 
\bold{B} = \left( \begin{array}{cc} 2 & 1 \\ 1 & 2 \end{array}\right) \;.
\ee 

The Kahler potential at the K-point, as an expansion in $s$, is
\begin{equation}
    e^{-K} = -\frac{3}{8\pi}\log\left[\frac{|s|^2}{3}\left(\frac{\Gamma\left(\frac13\right)^2}{2\pi}\right)^{12}\right]+\mathcal{O}(|s|)\;,
\end{equation}
which yields the corresponding moduli space metric
\begin{equation}
g_{s\bar{s}} \underset{s\rightarrow 0}{\sim} \frac{1}{4|s|^2\log(|s|)^2}\;.
\end{equation}
This matches expression (\ref{kinztos}).

\section{A quick review of Seiberg-Witten theory}
\label{sec:revSW}

In this appendix we present a quick review of Seiberg-Witten theory \cite{Seiberg:1994rs}, which is just $SU(2)$ pure gauge theory with ${\cal N}=2$ supersymmetry in four dimensions. We refer to \cite{Lerche:1996xu,Tachikawa:2013kta} for nice reviews. 

We consider pure $SU(2)$ ${\cal N}=2$ gauge theory. In the infrared, the theory has two types of frames or descriptions, depending on the value of the scalars in the vector multiplets. Let us parameterize the Coulomb branch by a complex scalar $a$. This is the parameterisation far away from the origin, in a sense that we make precise below. On the Coulomb branch, the gauge group is broken to the Cartan subgroup $SU(2) \rightarrow U(1)_E$. The subscript $U(1)_E$ denotes that this is the electric $U(1)$. The breaking scale is set by the value of $a$, and below that scale the gauge coupling of $U(1)_E$ stops running because there are no more charged states under it. So whether the infrared of the theory is weakly or strongly coupled can be parameterized in terms of the gauge coupling of the Cartan $U(1)_E$ inside $SU(2)$. We denote this complexified coupling by $S$, with 
\be 
S = \frac{\theta}{2\pi} + \frac{4 \pi i}{g^2} \;, 
\ee 
where $g$ is the gauge coupling, and $\theta$ the superpartner axion. Above the $SU(2)$ breaking scale this coupling runs, as part of the $SU(2)$. Because $a$ sets the mass of the W-bosons, we can write the coupling in the infrared as
\be
S = 2S_{UV}-\frac{2b}{2\pi i} \log \left(\frac{a}{\Lambda_{UV}}\right) + ...
\label{Sswexp}
\ee
where $b$ is, in general, the value of the $\beta$ function coefficient. For the case of pure $SU(2)$, we have $b=4$. $\Lambda_{UV}$ is the ultraviolet cutoff of the theory, and $S_{UV}$ is the value of the coupling at that scale of the $SU(2)$ gauge field (which differs by a factor of $2$ from that of the Cartan $U(1)$).

We can introduce a (complexified) dynamical scale $\Lambda$ given by 
\be 
\Lambda^4 = \left(\Lambda_{UV}\right)^4 e^{2\pi i S_{UV}} \;,
\ee 
and then write (\ref{Sswexp}) as
\be
S = -\frac{8}{2\pi i} \log \left(\frac{a}{\Lambda}\right) + ...
\ee
Therefore, the infrared descriptions of the theory are determined by the ratio $\frac{a}{\Lambda}$. We have
\bea 
\left|\frac{a}{\Lambda}\right| \gg 1 \;\;&:&\;\; \text{Weakly-coupled infrared} \;\;,\; \nn \\
\left|\frac{a}{\Lambda}\right| \ll 1 \;\;&:&\;\; \text{Strongly-coupled infrared} \;\;.
\label{SWphases}
\eea 

In the case of the strongly-coupled infrared, the Coulomb branch in the infrared is parameterized by a dual variable $a_D$, which is given in terms of the prepotential $F(a)$ as
\be 
a_D = \frac{\partial F(a)}{\partial a} \;,
\ee 
and at leading order we have
\bea 
F(a) &=& -\frac{4}{2\pi i} a^2 \log \left(\frac{a}{\Lambda}\right) + ... \;, \nn \\
a_D &=& -\frac{8}{2\pi i} a \log \left(\frac{a}{\Lambda}\right) + ... \;.
\eea
The theory is controlled by 
\be 
\left|\frac{a_D}{\Lambda}\right| \ll 1 \;.
\label{magreg}
\ee 
The $SU(2)$ is still broken, but now the theory should be described by a magnetic $U(1)_M$, so in the infrared we consider $SU(2) \rightarrow U(1)_M$. The magnetic gauge coupling $S_D$, in the infrared, is given by (at leading order)
\be 
S_D =\frac{1}{2\pi i} \log \left( \frac{a_D}{\Lambda} \right)+ ... \;.
\label{mcosw}
\ee 
So the magnetic gauge coupling is small in the strongly-coupled infrared phase. Further, (\ref{mcosw}) is exactly the form of a gauge coupling where a (BPS) state of mass $|a_D|$ and charge one under $U(1)_M$ has been integrated out. This state is the monopole state of Seiberg-Witten theory. 

In terms of the magnetic variable $a_D$, we can write the prepotential as
\be 
F\left(a_D\right) = \frac{1}{4\pi i} a_D^2 \log \left( \frac{a_D}{\Lambda} \right)+ ... \;.
\label{Fswadtemy}
\ee 
Then if we write it in terms of $S_D$, we have the exponential dependence
\be 
F\left(S_D\right) = \frac{1}{2} \;S_D \;e^{4 \pi i S_D}+ ... \;.
\label{expFdSWex}
\ee 
This exponential form can be compared to the expressions (\ref{ConifoldPrepogauge}) and (\ref{KPrepogauge}). In the case of Seiberg-Witten theory, and in the case of the conifold, we can understand it as arising from integrating out a light BPS monopole state.

\bibliographystyle{jhep}
\bibliography{Higuchi}

\end{document}